\documentclass[twocolumn,notitlepage,prx,superscriptaddress,longbibliography]{revtex4-2}
\usepackage{amsfonts}
\usepackage{textcomp}
\usepackage{times}
\usepackage{graphicx}
\usepackage{float}
\usepackage{latexsym,amsmath,amssymb,bm,euscript}
\usepackage{color}
\usepackage{subfigure}
\usepackage{epstopdf}
\usepackage[colorlinks=true,linkcolor=blue,citecolor=blue]{hyperref}
\usepackage{hyperref}
\usepackage{soul}
\usepackage[normalem]{ulem}
\usepackage{mathrsfs}
\usepackage{amsmath}
\usepackage{xspace}
\usepackage{natbib}
\usepackage{ulem}
\usepackage{etoolbox}
\usepackage{tikz}
\usepackage{physics}

\newcommand{\bq}{\mathbf{q}}
\newcommand{\bp}{\mathbf{p}}

\newcommand{\br}{\mathbf{r}}

\newcommand{\bL}{\mathbf{L}}
\newcommand{\bK}{\mathbf{K}}
\newcommand{\bR}{\mathbf{R}}
\newcommand{\bG}{\mathbf{G}}

\def\ket#1{|{#1}\rangle}

\begin{document}

\title{Fractional Chern Insulator and Quantum Anomalous Hall Crystal in Twisted MoTe$_2$}

\author{Jialin Chen}
\thanks{These authors contributed equally to this work.}
\affiliation{Institute of 
	Theoretical Physics, Chinese Academy of Sciences, Beijing 100190, China}

\author{Qiaoyi Li}
\thanks{These authors contributed equally to this work.} 
\affiliation{Institute of 
	Theoretical Physics, Chinese Academy of Sciences, Beijing 100190, China}
\affiliation{School of Physical Sciences, University of Chinese Academy of Sciences, Beijing 100049, China}

\author{Xiaoyu Wang}
\email{xywangmn@gmail.com}
\affiliation{National High Magnetic Field Lab, Tallahassee, Florida 32310, USA}

\author{Wei Li}
\email{w.li@itp.ac.cn}
\affiliation{Institute of Theoretical Physics, 
	Chinese Academy of Sciences, Beijing 100190, China}
 
\begin{abstract}
Recent experimental advances have uncovered fractional Chern insulator (FCI) states in twisted MoTe$_2$ (tMoTe$_2$) systems under zero magnetic field. Understanding the interaction effects on topological phases within realistic model presents a significant theoretical challenge. Here, we construct a moir\'e superlattice model tailored for tMoTe$_2$ and conduct investigations using state-of-the-art tensor-network methods. Our ground-state calculations reveal a rich variety of interaction-driven and filling-dependent topological phases, including FCIs, Chern insulators, and generalized Wigner crystals, which are revealed in recent experiments. For FCI state, dynamical simulations uncover a single-particle excitation continuum with a finite charge gap, reflecting the fractionalized charge excitations. Finite-temperature calculations further determine characteristic charge activation and ferromagnetic transition temperatures, reconciling existing experimental discrepancies. Furthermore, using this realistic lattice model, we predict the presence of quantum anomalous Hall crystals exhibiting integer Hall conductivity at fractional fillings in tMoTe$_2$. By integrating ground-state, finite-temperature, and dynamical analyses, our work establishes a comprehensive framework for understanding correlated topological phases in tMoTe$_2$ and related moir\'e systems.	
\end{abstract}

\maketitle

\section{Introduction}

Twisted bilayer transition metal dichalcogenides, such as twisted MoTe$_2$ (tMoTe$_2$), have
attracted significant research interest due to the emergence of correlated topological states \cite{Regnault2011Fractional,Neupert2011Fractional,Tang2011High,Sheng2011Fractional,Sun2011Nearly,
	Wu2019Topological,Zeng2023Thermodynamic,Cai2023Signatures,Xu2023Observation,Park2023Observation,
	Anderson2024Trion,Redekop2024Direct,Ji2024Local}. The observation of fractional Chern insulator (FCI)
~\cite{Regnault2011Fractional,Neupert2011Fractional,Tang2011High,Sheng2011Fractional,Sun2011Nearly}
represents a hallmark that has long been sought in condensed matter physics. Notably, FCI states at
filling factors $\nu_{\mathrm{h}}= -2/3$, $-3/5$, $-4/7$, $-4/9$, and $-5/9$ have been experimentally observed recently
in tMoTe$_2$ at zero magnetic field~\cite{Zeng2023Thermodynamic,Cai2023Signatures,Xu2023Observation,
	Park2023Observation,Anderson2024Trion,Redekop2024Direct}. It not only advances the understandings
of correlated topological quantum matter, but also shed light on fault-tolerant topological quantum computation \cite{Liu2025Fractional}.
Recent experiments have revealed signatures of (fractional) excitations in FCIs,
including conducting edge states~\cite{Ji2024Local}, charge gaps \cite{Redekop2024Direct}, and thermal
activation gaps~\cite{Park2023Observation}. Consequently, the exotic quantum states in tMoTe$_2$ ---
particularly correlated topological phases such as FCI and their generalizations --- have attracted significant
theoretical attention ~\cite{Li2021Spontaneous,kwan2024abelian,Crepel2023Anomalous,Reddy2023Toward,
	Wang2024Fractional,Reddy2023Fractional,Abouelkomsan2024Band,shen2024stabilizing,zaklama2024structure,
	Cheng2024Maximally,Sheng2024Quantum,Goldman2023Zero,wang2024interacting,wang2023topological,
	Liu2024Gate,Qiu2023Interaction,Yu2024Fractional,Dong2023Composite,Pan2022Topological}.

While charge orders typically compete with FCIs, we highlight that their intricate interplay can also stabilize novel quantum phases of matter. Prominent examples include quantum Hall crystals (QHCs) in strong magnetic fields~\cite{Halperin1986Compatibility, Kivelson1986Cooperative, Kivelson1987Cooperative, Teifmmode1989Hall, Murthy2000Hall} and, more remarkably, quantum anomalous Hall crystals (QAHCs) that emerge even at zero field~\cite{Song2024Intertwined}. Both states exhibit an integer quantized Hall conductivity despite a fractional electron filling. QAHCs further distinguish themselves by simultaneously breaking time-reversal, space group, and lattice translation symmetries.  While examples of QHCs and QAHCs have been observed in multilayer graphene systems~\cite{Su2025moire, lu2024extended, aronson2024displacement, waters2024interplay, zhang2024commensurate}, the microscopic mechanisms that stabilize these exotic states -- such as electron correlations and symmetry-breaking patterns -- are under active investigations~\cite{Trithep2024The, Sheng2024Quantum, Pan2022Topological, Lu_2024}. Indeed, the term “QAHC” is used here in a generalized sense, following its broader usage in recent studies of moir\'e systems. This usage differs from the original definition, which referred to spontaneous crystalline quantum anomalous Hall states in systems without an underlying periodic potential.

In particular, the exploration of QAHCs in semiconductor moir\'e systems has garnered considerable attention,
with theoretical proposals predicting their emergence in tMoTe$_2$~\cite{Sheng2024Quantum} and AB-stacked
MoTe$_2$/WSe$_2$ heterobilayers~\cite{Pan2022Topological}. Nevertheless, elucidating these correlated
topological states and studying their phase transitions presents substantial theoretical difficulties, owing to their
highly entangled many-body character. Exact diagonalization (ED) are limited to small clusters ($\sim$30
sites)~\cite{Sheng2024Quantum}, and Hartree-Fock methods fail to capture highly entangled phases like FCIs
\cite{Pan2022Topological}. More importantly, although a wide range of experiment measurements have been
performed, including spectroscopy~\cite{Cai2023Signatures,Ji2024Local,Anderson2024Trion,wang2025hidden},
transport~\cite{Xu2023Observation,Park2023Observation}, thermodynamics~\cite{Zeng2023Thermodynamic}, magnetometry~\cite{Redekop2024Direct}, and microstructural characterization~\cite{Li2021Imaging}, prior
theoretical efforts have mainly focused on the ground-state properties, leaving a critical gap between theory and
experiments~\cite{Trithep2024The}.

\begin{figure}[!tbp]
	\includegraphics[width=1\linewidth]{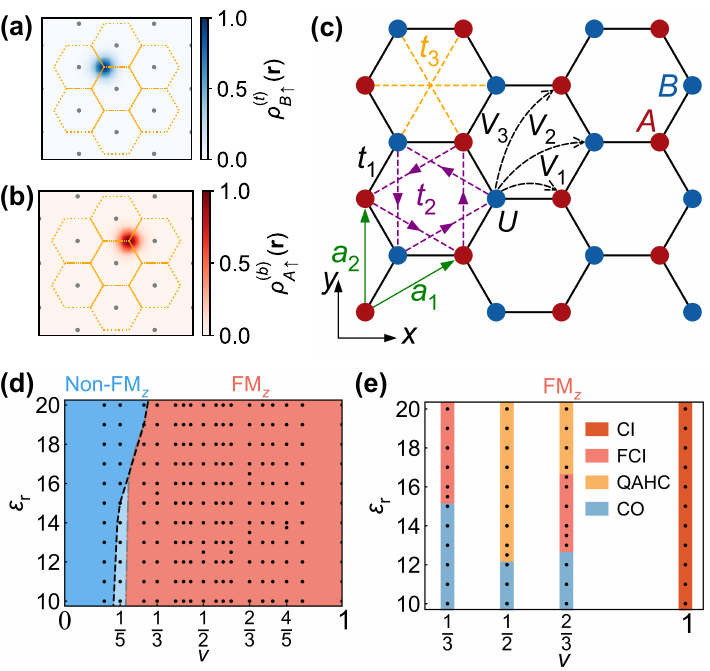}
	\caption{Real-space model on the honeycomb lattice and its global phase diagram. Panels (a) and (b)
		show real-space electronic density profile of exponentially localized Wannier orbitals, normalized
		to the peak density. $\rho^{(l)}_{i\uparrow} = \big| w^{(l)}_{i\uparrow} \big|^2$ for spin-$\uparrow$
		is centered on site $i=(A,B)$ in layer $l=(b,t)$. The Wannier orbitals reside mostly at the honeycomb
		lattice sites, corresponding to the XM/MX stacking regions of the twisted bilayer MoTe$_2$.
		(c) Schematic representation of the model with hopping parameters ($t_1$, $t_2$, $t_3$)
		and interactions ($U$, $V_1$, $V_2$, $V_3$) up to the third nearest neighbor on a honeycomb
		lattice. $\boldsymbol{a}_1=(\sqrt{3}/2, 1/2)$ and $\boldsymbol{a}_2=(0,1)$ denote the primitive
		vectors of the honeycomb lattice. Numerical simulations employ a X-cylinder (XC) geometry
		with periodic boundary conditions along $\boldsymbol{a}_2$ direction.
		(d) Magnetic phase diagram in the $\epsilon_\mathrm{r}$-$\nu$ plane, where the dashed line (obtained
		with $N_y=5$ cylinder) and dash-dotted boundary lines ($N_y=4$ data) separate the FM$_z$ and
		non-FM$_z$ phases. (e) Phase diagrams with representative fillings ($\nu = 1, 2/3, 1/2$,
		and $1/3$) contains various phases, including the CI, FCI, QAHC, and CO phases. The phase
		diagram is constructed using width-6 ($N_y=6$) simulations, a system size that accommodates the
		root configurations of both FCIs and composite Fermi liquids~\cite{Haldane1991statistics,Scott2016The}. }
	\label{Fig1}
\end{figure}

To address these fundamental questions and bridge the existing gap, we first construct a realistic Hubbard-type model for tMoTe$_2$ in real space (see Fig.~\ref{Fig1}), constructed from properly chosen Wannier orbitals. This real-space approach provides a distinct perspective from prevailing numerical studies of moir\'e systems, which predominantly employ momentum-space~\cite{Li2021Spontaneous, kwan2024abelian, Crepel2023Anomalous, Reddy2023Toward,Wang2024Fractional, Reddy2023Fractional, Abouelkomsan2024Band, shen2024stabilizing, zaklama2024structure, Cheng2024Maximally, Sheng2024Quantum, Goldman2023Zero} or hybrid real-space/momentum-space representations~\cite{Soejima2020Efficient, Parker2021Strain, Wang2023Ground, Dong2023Composite}. We then employ comprehensive, unbiased, large-scale tensor-network calculations to investigate the ground-state, finite-temperature, and dynamical properties, connecting our results with experimental findings on tMoTe$_2$.

In our ground-state analysis, we identify the emergence of ferromagnetic (FM) order and FCI states across electron fillings $1/3 \leq \nu <1$, in agreement with experimental observations, while also revealing a competitive interplay between CO and FCI phases. Through a particle-hole transformation --- detailed further below --- this filling range corresponds to hole fillings of $-1 < \nu_h \leq -1/3$. Remarkably, QAHC states are revealed at $\nu = 1/2, 2/3, 3/5, 4/5$ and $0.63$, arising from two distinct mechanisms --- hole crystal formation and band folding. We further propose that these QAHC states could be driven into FCI states by melting their charge order through tuning parameters like dielectric constant and temperature.

Our finite-temperature calculations further reveal a non-monotonic behavior of FM transition temperature $T_c$ versus fillings, in agreement with experimental observations~\cite{Cai2023Signatures}. Furthermore, we also compute the single-particle spectral function at zero temperature, for several representative fillings including $\nu = 1/2$, $2/3 $, and $1$, and identify the charge gaps as approximately 18.2, 8.4, and 6.2 meV, respectively. Finite-temperature simulations further yield results of the local density of state at the Fermi level $ A_{\rm loc}(\omega = 0)$, which disclose a typical thermal activation temperature of about $T^* \approx 30$~K
for $\nu = 2/3, 1$.  Notably, the zero-temperature charge gaps are found to be significantly larger than the energy scales of thermal activation and FM transition, a phenomenon that has been also observed in experiments but lacks theoretical understanding~\cite{Redekop2024Direct, Park2023Observation}. Our numerical results reproduce these experimental observations and resolve their contradictions with a realistic real-space model, which establish a robust foundation for further theoretical and experimental investigations on MoTe$_2$ and moir\'e materials in general.

\section{Real-space model on the honeycomb lattice}

Bilayer MoTe$_2$ with a twist angle of approximately 3.89$^\circ$ has recently attracted great research interest
\cite{Zeng2023Thermodynamic,Park2023Observation,Xu2023Observation,Cai2023Signatures,Redekop2024Direct,
	Anderson2024Trion}. To provide a theoretical understanding, density functional theory (DFT) calculations \cite{Wu2019Topological,wang2023topology,Wang2024Fractional,Reddy2023Fractional,WuQuanSheng2024} have been
conducted to reveal the relevant low-energy degrees of freedom. They comprise two bands per valley with Chern numbers
$+1$ and $-1$, respectively. Due to the spin-valley locking \cite{DiXiao2012}, the spin-$\uparrow$ state is associated with
valley $\bK$, while the spin-$\downarrow$ state is linked to valley $\bK'$, resulting in a minimal model encompassing
four electronic bands.
For simplicity, we refer primarily to the spin states throughout the manuscript, omitting explicit
reference to the valley degree of freedom. It has been shown that the low energy physics of tMoTe$_2$ can be effectively
described by a continuum model \cite{Wu2019Topological} (see also Supplementary section~I online), which can
be mapped to a tight-binding model through a proper choice of Wannier basis, through the trial state procedure
\cite{Soluyanov2011Wannier,Qiu2023Interaction,Cheng2024Maximally}. As illustrated in Figs.~\ref{Fig1}a
and b, the Wannier orbitals $w^{(l)}_{i\uparrow}$ for spin-$\uparrow$ are exponentially localized on the two
sublattices ($i=A$ or $B$) of a honeycomb lattice~\cite{Soluyanov2011Wannier} and exhibit clear layer resolution
($l=b$ or $t$, meaning bottom or top). Specifically, $w^{(b)}_{A\uparrow}$ predominantly resides on the bottom layer,
while $w^{(t)}_{B\uparrow}$ resides on the top layer. The situation is the same for spin-$\downarrow$ orbitals since
the two spin orientations are related by the time-reversal symmetry.

By projecting the kinetic term and Coulomb interactions onto these Wannier orbitals, we derive a Hubbard-type
real-space model for tMoTe$_2$ on the honeycomb lattice, as depicted in Fig.~\ref{Fig1}c. The Hamiltonian reads
\begin{equation}{\label{Eq_Model}}
	\begin{split}
		H = & \sum_{(i, j)} \big( t_{ij } c^\dagger_{i\uparrow} c_{j\uparrow} + t^*_{ij } c^\dagger_{i\downarrow} c_{j,\downarrow} + h.c. \big)
		-\mu \sum_i  n_{i}  \\ & + U\sum_{i} n_{i\uparrow} n_{i\downarrow }+  \sum_{  i<j } V_{ij} n_i n_j,
	\end{split}
\end{equation}
where $c^\dagger_{i\sigma}$ ($c_{i\sigma}$) creates (annihilates) an electron with spin $\sigma=\uparrow$, $\downarrow$ on site $i$. The density at site $i$ is $n_i = n_{i\uparrow} + n_{i\downarrow}$, with $n_{i\sigma} =c^\dagger_{i\sigma} c_{i\sigma}$, and $\mu$ is the chemical potential. The hopping $t_{ij}$ is defined for each $(i,j)$ bond in the spin-$\uparrow$ sector, with the corresponding terms for spin-$\downarrow$ obtained by Hermitian conjugation. We include hopping amplitudes up to the third nearest neighbor (\(t_1, t_2, t_3\)) in our model (see Fig.~\ref{Fig1}c). Notably, the next-nearest neighbor hopping $t_2$ for spin-$\uparrow$ electrons along the direction depicted in Fig.~\ref{Fig1}c incorporates a phase factor of $\phi = 2\pi/3$. Therefore, hopping parameters (for spin-$\uparrow$) are $t_1 = -3.225$~meV, $t_2 = 2.120 e^{i2\pi/3}$ meV, and $t_3 = 0.947$~meV.

The onsite Coulomb interaction is denoted as $U = 1192.71/\epsilon_\mathrm{r}$~meV, where $\epsilon_\mathrm{r}$ represents the
relative dielectric constant (see Supplementary section~I online). For further couplings $V_{ij}$, we consider a dual-gate screened Coulomb interaction in tMoTe$_2$, i.e.,
\begin{equation}
	V(r) =  \frac{e^2}{4\pi r\epsilon_0\epsilon_\mathrm{r}}e^{-r/\xi}  = \frac{276.92}{\epsilon_\mathrm{r}  r/L_m} \exp\Big(-\frac{r/L_m}{\xi/L_m} \Big)~\text{meV},
\end{equation}
where $\epsilon_0$ is the vacuum permittivity, $r$ is the distance between the two sites (i,j), and $\xi$ is the screening length. $L_m \simeq 5.2$~nm represents the length of primitive vectors of the moir\'e superlattice (see Fig.~\ref{Fig1}c), which is taken as the unit of length in the present study. In our simulations, we mainly consider $10 \leq \epsilon_\mathrm{r} \leq 20$ in tMoTe$_2$, aligning with previous studies~\cite{Wang2024Fractional,wang2024interacting,Cheng2024Maximally}. In practice, we adopt a reasonable value of $\xi/L_m = 2$, as also used in Ref.~\cite{Cheng2024Maximally}, and retain interaction terms up to the third nearest neighbor (i.e., $V_1, V_2, V_3$ as indicated in Fig.~\ref{Fig1}c), compatible with our finite width of cylinder (up to 6). 
Despite the larger experimental $\xi \simeq 30~{\rm nm} (\simeq 5.77 L_m)$, this truncation in $V_{ij}$ and $t_{ij}$ remains a valid approximation, supported by analysis in Supplementary Section~I.

In this study, we investigate the zero-temperature~\cite{White1992Density, mcculloch2008infinite, SCHOLLWOCK2011The, tenpy2024}, finite-temperature~\cite{Li2023Tangent}, and dynamical~\cite{Haegeman2011PRLTime, Haegeman2016PRBUnifying, Kuehner1999PRBDynamical, Li2022PRRDetecting, Gleis2023PRLControlled, Li2024PRLTime} properties using state-of-the-art tensor-network methods (see Supplementary section~II online). The simulations are conducted on an X-cylindrical (XC) geometry of size $N_y \times N_x \times 2$. Here for the sake of simplicity, we have performed a particle-hole transformation in the model~(\ref{Eq_Model}). It inverts the hopping parameters, chemical potential, and thus the non-interacting band structure~\cite{Reddy2023Fractional} (see also Supplementary section~I online). In the following, we use $\nu$ to denote the electron filling per unit cell, which consists of two lattice sites, $A$ and $B$. Consequently, the experimentally relevant hole filling $-1 \leq \nu_h \leq 0$ thus corresponds to the electron filling $0 \leq \nu \leq 1$ in our practical model calculations. Moreover, to reduce computational costs, the Hubbard-type model described by Eq.~(\ref{Eq_Model}) can be simplified to either a spinful or spinless $t$-$V$ model (Supplementary section~II online).

\section{Ground-state results}
\subsection{Global phase diagram}

The presence of FM order has been experimentally identified as a prerequisite for the emergence of FCI and other correlated
topological states in MoTe$_2$~\cite{Zeng2023Thermodynamic, Park2023Observation, Xu2023Observation, Cai2023Signatures, Redekop2024Direct, Anderson2024Trion, Ji2024Local}. Therefore, we first examine the emergence of FM order, characterized by full polarization of spin-$S_z$ (denoted as FM$_z$). To evaluate the ground-state magnetism across various filling factors, we compare the energies of different $S_z$ sectors in our iDMRG simulations (see Supplementary section~III online). Our results reveal a finite energy gap above the FM$_z$ ground state for $\nu \gtrsim 0.2$, as indicated by darker blue shading in Fig.~\ref{Fig1}{d} for $N_y=4$ and $5$. Additionally, for $N_y=6$, we find that FM$_z$ order remains robust at filling factors $\nu=1$, $2/3$, $1/2$, and $1/3$, consistent with recent experimental observations~\cite{Zeng2023Thermodynamic, Cai2023Signatures}.

To investigate the presence of quantized Hall conductivity in the FM$_z$ phases, we employ a charge pumping
method \cite{Laughlin1981Quantized,Grushin2015Characterization,Chen2021Realization} within our iDMRG
simulations. For filling factors $\nu=p/q$, we
adiabatically introduce a flux of $\Phi_y = 2q\pi$ along the cylinder axis by modulating the hopping terms
with a phase factor $e^{i\Phi_y}c^\dagger_{i\sigma}c_{j\sigma} + h.c.$ as the electron traverses the periodic
boundary. For a charge pumping of $\Delta Q$, the corresponding Hall conductivity is given by $\sigma_{xy}
= \frac{2\pi|\Delta Q|}{\Phi_y}$, whose unit is set as $e^2/h$ in this study.

Our simulations, primarily conducted for fillings $\nu=p/q$ and $q=1,2,3$, reveal the existence of quantized Hall
conductivity $\sigma_{xy}$, as indicated in Fig.~\ref{Fig1}{e}. We also consider $q=5,7,9$ cases on $N_y=4, 5$
cylinders, with the results summarized in Fig.~S8 (Supplementary section~III online). Notably, we successfully observe
experimentally relevant $\sigma_{xy}=\nu=1$ for CI states (Supplementary section~III online) and $\sigma_{xy}=\nu=2/3$, $4/7$, $5/9$,
$4/9$, $3/5$ for FCIs. Intriguingly, we predict a robust FCI state with $\sigma_{xy} = \nu = 1/3$ and advocate for
its experimental observation in future studies. Moreover, we also identify instances where an integer Hall conductivity
of $\sigma_{xy} = 1$ can be observed at fractional fillings $\nu = 1/2$, $2/3$ (for $N_y = 6$ cylinder), as well as
at $\nu = 3/5$ and $4/5$ (for $N_y = 5$). An even more exotic QAHC state emerges at an incommensurate filling $\nu \simeq 0.63$, as shown in Fig.~S9~(Supplementary section~III online). These integer Hall states coexist with COs that are compatible with the
cylinder geometry, leading to the formation of QAHC states.

\begin{figure*}[!htbp]
	\includegraphics[width=1\linewidth]{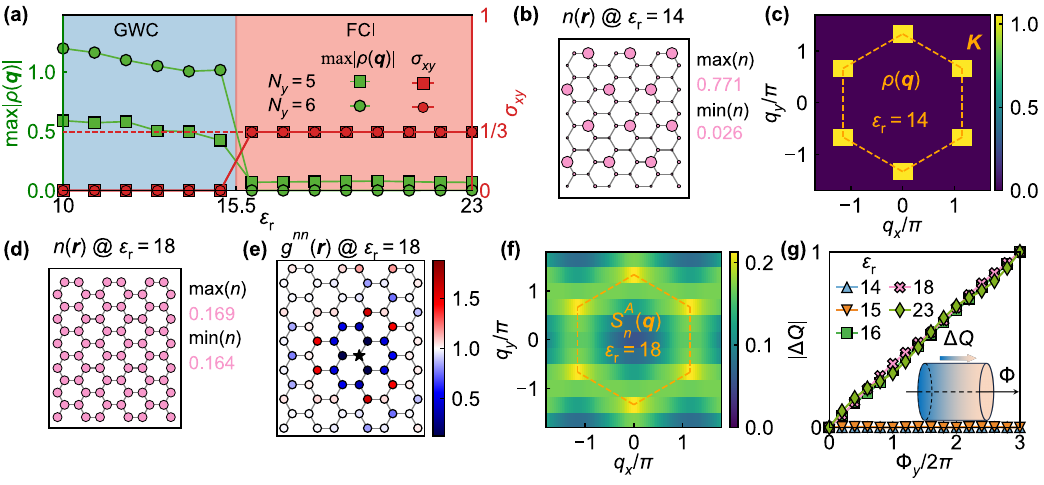}
	\caption{GWC and FCI phases for $\nu=1/3$ filling.
		(a) Phase diagram containing the GWC and FCI states tuned by the relative dielectric constant $\epsilon_\mathrm{r}$.
		The two phases are characterized by $\rho({\boldsymbol{q}})$ and the Hall conductivity $\sigma_{xy}$, obtained on
		$N_y=5$ and $6$ cylinders.Charge distribution (b) $n(\boldsymbol{r})$ in real-space  and (c) $\rho(\boldsymbol{q})$ in momentum space for the GWC state (with $\epsilon_\mathrm{r}=14$), showcasing the triangular lattice pattern of electron density modulation.  (d) demonstrates $n(\boldsymbol{r})$ for the FCI state ($\epsilon_\mathrm{r}=18$), where a uniform charge distribution is observed. 		
		(e) Charge-charge correlation function $g^{nn}(\boldsymbol{r})$ in the FCI phase
		(with $\epsilon_\mathrm{r}=18$), displaying the charge distribution relative to the reference site $n_0$ marked by a black star.
		(f) Charge structure factor $S^A_n(\boldsymbol{q})$ in momentum space of the FCI state.
		(g) Charge pumping results $|\Delta Q|$ for various $\epsilon_\mathrm{r}$ for $N_y=6$, measured after inserting a flux
		$\Phi_y=6\pi$. The inset illustrates the charge pumping simulation process. For this and subsequent plots, the dashed
		hexagon [c.f., panels (c) and (f)] outlines the first Brillouin zones of the honeycomb lattice.
	}
	\label{Fig_nu13_2by4}
\end{figure*}

Regarding the COs, we have thoroughly examined the charge density distribution $n(\boldsymbol{r})$
in real space and $\rho(\boldsymbol{q}) = \frac{1}{\sqrt{N}} \big | \sum_{i} e^{i\boldsymbol{q} \cdot
	\boldsymbol{r}_i} (\langle n_i \rangle - \bar{n})\big|$ in momentum space, where $\boldsymbol{r}_i$
labels the position of a unit cell (rather than the individual A/B site), revealing the existence of CO
states. As illustrated in Fig.~\ref{Fig1}{e} (highlighted in blue), these topologically trivial CO
states are located in the lower-left corner of the phase diagram, characterized by low fillings and
strong interactions. The nontrivial interplay between COs and topological phases will be explored
in subsequent sections through representative examples.

\subsection{GWC and FCI for $\nu=1/3$}

To start, we consider the filling fraction $\nu = 1/3$, and present the results on $N_y = 5$ and $6$
cylinders. Our results reveal that the
ground states exhibit FM$_z$ order for a broad range of $\epsilon_\mathrm{r}$ values, specifically $10 \leq
\epsilon_\mathrm{r} \leq 23$. For the regime with a relatively small $\epsilon_\mathrm{r}$, a GWC phase appear in the
phase diagram Fig.~\ref{Fig_nu13_2by4}{a}. The real-space charge density $n(\boldsymbol{r})$
is centered at the $A$ sublattice and forms a triangular lattice with primitive vectors
$\boldsymbol{a}^t_1=(\sqrt{3}, 0)$ and $\boldsymbol{a}^t_2 =(\sqrt{3}/2,3/2)$, thus breaking the
translational symmetry of the original honeycomb lattice (see Fig.~\ref{Fig_nu13_2by4}b).
Correspondingly, in Fig.~\ref{Fig_nu13_2by4}c we find clear peaks in $\rho(\boldsymbol{q})$
at the vertices ($\boldsymbol{K}$ points), namely, the characteristic reciprocal vectors
$\boldsymbol{b}^t_1=(0, 4\pi/3)$ and $\boldsymbol{b}^t_2 =(-2\sqrt{3}\pi/3, 2\pi/3)$ of the triangular
lattice. The results confirm the presence of a triangular GWC for relatively small $\epsilon_\mathrm{r}$.

However, for a larger relative dielectric constant, e.g., $\epsilon_\mathrm{r} = 18$, the charge density
$n(\boldsymbol{r})$ becomes uniformly distributed across the honeycomb lattice (Fig.~\ref{Fig_nu13_2by4}d). As a result, $\rho(\boldsymbol{q})$ becomes negligible
across the entire momentum space (Fig.~\ref{Fig_nu13_2by4}{a}). As seen in Fig.~\ref{Fig_nu13_2by4}{a}, the maximum of $\rho(\boldsymbol{q})$ undergoes a sharp
drop to very small values at $\epsilon^{c}_r=15.5$. For even larger $\epsilon_\mathrm{r} > 23$,
the ground state no longer has an FM$_z$ order.

To characterize the transitions between GWC and FCI, we compute the charge-charge correlations
$g^{nn}(\boldsymbol{r}) = \frac{\langle n_0 n_i\rangle}{\langle n_0\rangle\langle n_i\rangle}$ between
the charge density $n_0$ at a fixed site and other charge densities $n_i$ located at site $i$. This
metric provides insights into the attractive ($g^{nn}>1$) and repulsive ($g^{nn}<1$) correlations
between electrons. As depicted in Fig.~\ref{Fig_nu13_2by4}{e} for $\epsilon_\mathrm{r}=18$, we find
$g^{nn} < 1$ for the first three nearest neighbors of the reference site, indicating a repulsive interaction
among electrons within this close proximity. However, $g^{nn}$ exceeds 1.0 for the fourth and fifth
nearest neighbors, suggesting an attractive correlation in this region. This oscillating pattern of repulsion
and attraction can extend even further away from the reference site, which constitutes a characteristic
of neutral collective excitation, roton, in a FCI liquid~\cite{Girvin1986Magneto-roton}.

We examine the charge structure factor $S^{A}_n(\boldsymbol{q})$ on the $A$ sublattice
\begin{equation}
	S^{A}_n(\boldsymbol{q})= \frac{1}{N} \Big| \sum_{i, j\in A}
	e^{i\boldsymbol{q}(\boldsymbol{r}_i - \boldsymbol{r}_j)}(\langle n_in_j\rangle - \langle n_i\rangle\langle n_j\rangle)\Big |,
\end{equation}
which quantifies the spatial distribution of charge correlations in momentum space. In Fig.~\ref{Fig_nu13_2by4}f,
we plot $S^A_n(\boldsymbol{q})$ for $\epsilon_\mathrm{r}=18$ in the uniform state with $\rho^m(\boldsymbol{q})\simeq 0$.
The $S^A_q(\boldsymbol{q})$ peaks at the $\boldsymbol{K}$ points, which  reflect the oscillating pattern
observed in $g^{nn}(\boldsymbol{r})$. The positions of the peaks align with those of the GWC peaks depicted
in Fig.~\ref{Fig_nu13_2by4}c, indicating that the condensation of roton modes may be responsible for
the transition from a uniform FCI to a triangular-lattice GWC.

Moreover, we simulate the Hall conductivity through charge pumping, as illustrated in Fig.~\ref{Fig_nu13_2by4}g
for several values of $\epsilon_\mathrm{r}$ and $N_y=6$, using iDMRG calculations. Our results demonstrate that inserting a flux
of $\Phi_y=6\pi$ induces a charge pumping of $|\Delta Q|=1$ (thus $\sigma_{xy}=1/3$) for FCIs, while $|\Delta Q|=0$ for
GWCs. By combining the Hall conductivity $\sigma_{xy}$ and the charge distribution results, we construct the phase
diagram in Fig.~\ref{Fig_nu13_2by4}{a}: For $10 \leq \epsilon_\mathrm{r} \leq 15.5$, the system resides in a topologically
trivial GWC phase with $\sigma_{xy}=0$; in contrast, within the range $15.5 < \epsilon_\mathrm{r} \leq 23$, the ground state
transitions to a uniform FCI state with a quantized Hall conductivity of $\sigma_{xy}=1/3$.

\subsection{Hole crystal for $\nu=2/3$}
\begin{figure}[h]
	\includegraphics[width=1\linewidth]{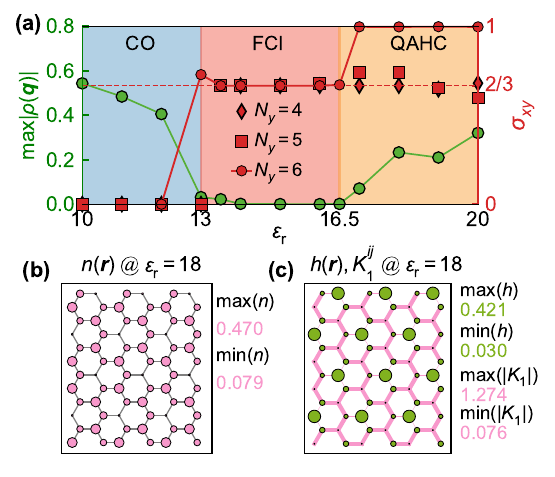}
	\caption{Emergence of QAHC states for $\nu=2/3$ filling.
		(a) Phase diagram containing CO, FCI, and QAHC phases versus $\epsilon_\mathrm{r}$, characterized by
		$\rho(\boldsymbol{q})$ and $\sigma_{xy}$. (b) Charge density $n(\boldsymbol{r})$ and (c)
		hole density $h(\boldsymbol{r})$ distributions in the QAHC state obtained with $\epsilon_\mathrm{r}=18$ and on
		$N_y=6$ cylinder. As we define an average density of $n=0.5$ (per site) as the ``full'' filling $\nu=1$ (per
		unit cell) of the CI phase, the hole filling is thus $h(\boldsymbol{r}) =0.5-n(\boldsymbol{r})$, as depicted
		by green dots. The pink lines in panel (c) show the absolute values of $K^{ij}_1$ (see definition in
		the main text).
	}
	\label{Fig_nu23_Crystal_3by4}
\end{figure}

In Fig.~\ref{Fig_nu23_Crystal_3by4}, we show the results for fractional filling $\nu=2/3$. Our scanning reveals
that the ground state exhibits FM$_z$ order for $\epsilon_\mathrm{r}$ betwee 10 and at least 30, for $N_y=4,5,6$
cylinders. In Fig.~\ref{Fig_nu23_Crystal_3by4}{a}, we present the results of $\rho(\boldsymbol{q})$ and
charge pumping for various $\epsilon_\mathrm{r}$ and widths. Notably, for a system width of $ N_y = 6 $, the system
transitions from a CO state at relatively small $ \epsilon_\mathrm{r} < 13 $, to a uniform FCI state with maximum of
$\rho(\boldsymbol{q}) \simeq 0 $ and $ \sigma_{xy} = 2/3 $ for $ 13 \leq \epsilon_\mathrm{r} \leq 16.5$.
Furthermore, for $\epsilon_\mathrm{r} > 16.5$ (and up to about $\epsilon_\mathrm{r} \approx 30$), a QAHC state emerges with
Hall conductivity of $\sigma_{xy} = 1$ and an underlying GWC order formed by holes (see Fig.~\ref{Fig_nu23_Crystal_3by4}).

We emphasize that the QAHC state exhibits strong geometric sensitivity. For the $N_y = 6$ cylinder, we can observe a quantized Hall conductivity of $\sigma_{xy} = 1$; however, such a state is absent for $N_y = 4$ and $N_y = 5$. This can be understood by examining the underlying charge distributions. In Fig.~\ref{Fig_nu23_Crystal_3by4}{b}, we show real-space charge distribution $n(\boldsymbol{r})$, where the maximum density $n \simeq 0.470$ and minimum density close to zero. In Fig.~\ref{Fig_nu23_Crystal_3by4}{c}, the holes form a triangular GWC structure with a
period of 3 along the width ($\mathbf{a}_2$) direction, which is compatible with the width $N_y=6$.

\begin{figure}[!bp]
	\includegraphics[width=1\linewidth]{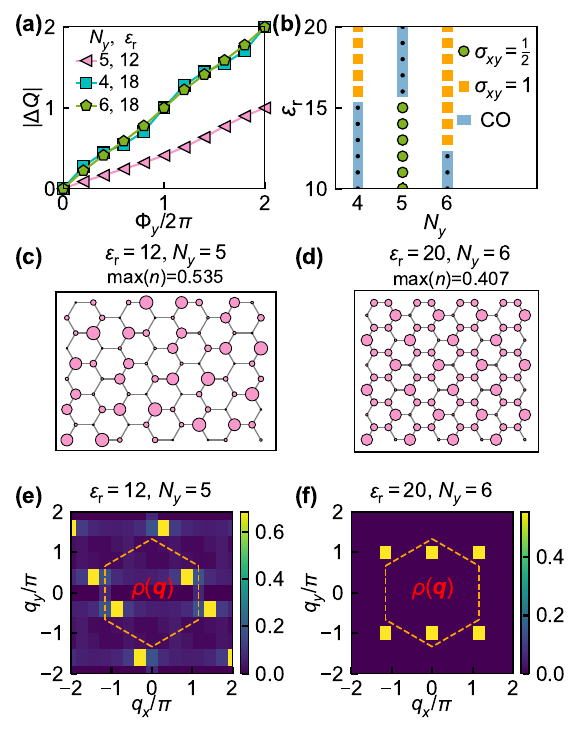}
	\caption{FCI and QAHC for $\nu=1/2$ filling. (a) Charge pumping $|\Delta Q|$ after inserting flux $\Phi_y=4\pi$, for FCI and QAHC states with different $\epsilon_\mathrm{r}$ and $N_y$. (b) The quantized $\sigma_{xy}$ results obtained from charge pumping for various $\epsilon_\mathrm{r}$ and $N_y$.  Panels (c,e) show $n(\boldsymbol{r})$ and $\rho(\boldsymbol{q})$, respectively, revealing non-uniform charge distribution for the FCI state obtained with $N_y=5$ and $\epsilon_\mathrm{r}=12$. Panels (d, f) show results of the QAHC state for $N_y=6$ and $\epsilon_\mathrm{r}=20$.
	}
	\label{Fig_nu12_D_3by4}
\end{figure}

Furthermore, in Fig.~\ref{Fig_nu23_Crystal_3by4}{c} we show the absolute values of nearest hopping $K^{ij}_1 = \left| \expval{t_1 \sum_{\sigma} (c^ \dagger_{i\sigma}c_{j\sigma} + h.c.)} \right |$, represented by the pink lines. Notably, the hopping between the largest green hole dots and other sites is virtually absent, suggesting that these holes are ``immobile" and isolated from the electron background. Consequently, the QAHC can be interpreted as a composite state comprising an CI state of electrons with $\nu=1$ (thus Hall conductivity $\sigma_{xy}=1$) and a GWC of holes with density $\nu_h=-1/3$. Such an QAHC is also named a hole crystal state \cite{Trithep2024The}. Moreover, as shown in Fig.~\ref{Fig_nu23_Crystal_3by4}{a} for $N_y = 4$ and 5, the FCI state (with possible
COs coexisting) exists, instead of QAHCs, due to the lattice geometry mismatches and has $\sigma_{xy}=2/3$. \\

\begin{figure*}[!htbp]
	\centering
	\includegraphics[width = \linewidth]{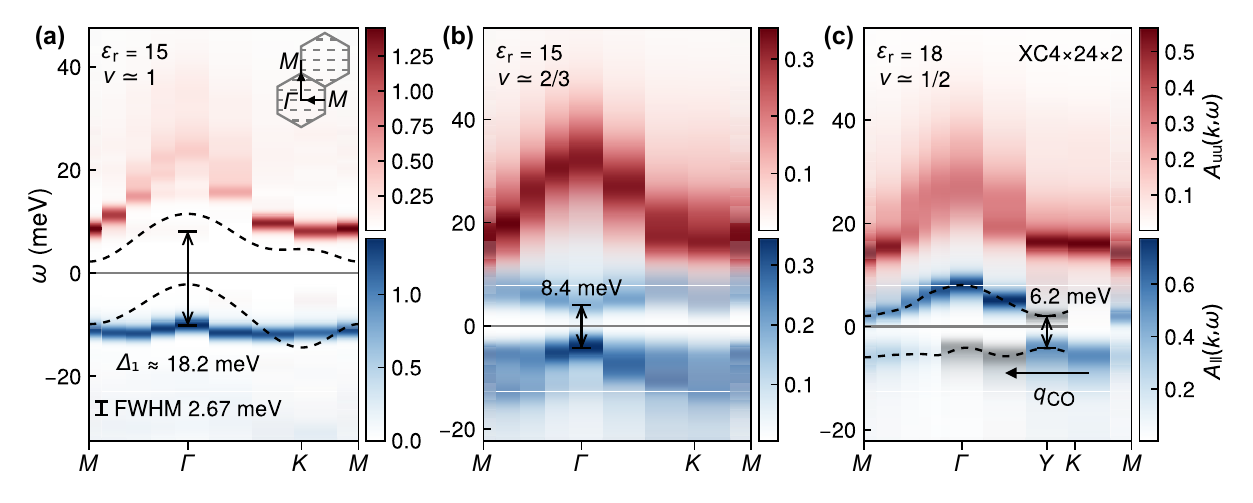}
	\caption{Charge excitation spectra in CI, FCI and QAHC phases.
		The band-resolved spectral functions $A_{\rm uu}(\bold{k}, \omega)$ and $A_{\rm ll}(\bold{k}, \omega)$
		simulated at zero temperature are shown for (a) CI state at $\nu = 1$, (b) FCI state at $\nu = 2/3$,
		and (c) the QAHC state at $\nu = 1/2$. The followed path in the reciprocal space is indicated in the inset
		of (a). As a comparison, the bare band dispersion (without interactions) is illustrated by the dashed lines
		in (a). The gray shadow in (c) represents the band folding with vector $\boldsymbol{q}_{\rm CO} =
		(0, -\pi)$. The peaks of the two folded bands are marked by dashed lines (serving as a guide for the eye).
		The frequency resolution is approximately 2.67~meV, as indicated by the range bar in (a).
	}
	\label{Fig_GSAkomega}
\end{figure*}

\subsection{QAHC with band folding for $\nu=1/2$ }
Besides the fractional fillings $\nu= p/q$ with odd denominators $q$, below we also consider an even-$q$ filling fraction $\nu = 1/2$ in this study. As shown in Fig.~\ref{Fig_nu12_D_3by4}a,
charge pumping simulations on systems with an odd width $N_y = 5$ (with $\epsilon_\mathrm{r} = 12$) find a fractional Hall conductivity $\sigma_{xy} = 1/2$, indicating the emergence of a FCI state. However, the same simulations on even-width systems with $N_y = 4, 6$ (with a different $\epsilon_\mathrm{r} = 18$, though) find an integer Hall conductivity $\sigma_{xy} = 1$, indicating the presence of QAHC state.

To understand such a width dependence, we compare the COs in the ground states. In Fig.~\ref{Fig_nu12_D_3by4}, we show that for $N_y = 5$, the charge distribution $n(\boldsymbol{r})$ is incommensurate with the honeycomb lattice and $\rho(\boldsymbol{q})$ shows peaks at $(0.866\pi, 0.4\pi)$ and $(-0.866\pi, -0.4\pi)$. In contrast, for $N_y = 6$, a well defined commensurate CO emerges, as evidenced by the $(0, \pm\pi)$ peaks of $\rho(\boldsymbol{q})$ in momentum space. For even $N_y$, the CO with an enlarged unit cell allows an integer filling of the folded band (see Fig.~\ref{Fig_GSAkomega} and related discussions below), yielding integer Hall conductivity. Note that a recent ED study has also reported a QAHC state at $\nu=1/2$ filling {but with a much larger symmetry-breaking pattern in real space}~\cite{Sheng2024Quantum}. Here our large-scale simulation results further reveal a sensitive even-odd width dependence in the ground state. The emergence of a commensurate CO in even $N_y$ systems, and its absence for odd $N_y$, may explain the observations for $\nu = 1/2$ filling. It offers an band-folding mechanism for the presence of QAHC, different from the hole crystal picture discussed above for $\nu = 2/3$. However, we emphasize that these two scenarios --- hole crystallization and band folding --- are not mutually exclusive.

\section{Excitation gap and charge continuum}
To study the dynamical properties, we compute the band-resolved spectral functions $A_{mm}(\bold{k}, \omega) \equiv -2{\rm Im} G_{mm}^R(\bold{k}, \omega)$ of the CI state at $\nu = 1$, the FCI state at $\nu = 2/3$, and the QAHC state at $\nu = 1/2$. $G_{mm}^R(\bold{k}, \omega)$ is the retarded Green's function and $m = {\rm u}, {\rm l}$ labels the upper/lower band (see Supplementary section~V online for details). The dynamical results reveal distinct features for these topological states, including the flattened band, charge excitation continuum, and band folding, providing valuable insights into their distinct nature.

Starting from the $\nu = 1$ CI phase in Fig.~\ref{Fig_GSAkomega}{a}, we find the lower band becomes significantly flattened compared to the bare band (with interactions turned off), highlighting
the correlation effect. Additionally, a band gap of approximately $\Delta_1 \approx 18.2 \, \text{meV}$ is obtained, with the Fermi level lying within the gap, consistent with the gapped CI phase at $\nu = 1$.
On the other hand, in Fig.~\ref{Fig_GSAkomega}{b}, regarding the FCI state with $\nu = 2/3$, both the upper and lower bands exhibit significant broadening. The charge excitation continuum, instead
of well-defined quasi-particle modes, signposts the emergence of fractional charge excitations in the FCI phase. Moreover, between the two broad ``sub-bands'', there exists a charge gap $\Delta_{2/3} \approx 8.4 \, \text{meV}$, where the Fermi level resides, supporting a gapped nature of FCI. The charge gaps at $\nu = 1$ and $\nu = 2/3$ align well with the experimental results of $14\pm 1$~meV and $7.0 \pm 0.5$~meV, respectively, as obtained by direct magnetic imaging at a low temperature of 1.6~K~\cite{Redekop2024Direct}.

Lastly, for the QAHC phase at $\nu = 1/2$, as shown in Fig.~\ref{Fig_GSAkomega}{c}, the lower band
also splits into two sub-bands, with a charge gap {$\Delta_{1/2}\approx 6.2$ meV}. Unlike the charge-uniform
FCI states, there exists shadow bands near, e.g., the $\boldsymbol{\Gamma}$ point in the lower sub-band.
This characteristic feature suggests the occurrence of Brillouin zone folding, due to the presence of CO in the
QAHC phase. Considering the CO wave vector $\boldsymbol{q}_{\rm CO} = (0, \pm \pi)$, as observed from
$\rho(\boldsymbol{q})$ in Fig.~\ref{Fig_nu12_D_3by4}{f}, we shift the spectra of the lower two
sub-bands. As seen in Fig.~\ref{Fig_GSAkomega}c, the spectral weights shift from the $Y$–$M$ segment
to $\Gamma$–$Y$, resulting in two folded bands. The Fermi level lies within the gap between these two
bands, with the lower folded band being significantly flatter than the bare band. This closely resembles the
band structure of the $\nu = 1$ CI state shown in Fig.~\ref{Fig_GSAkomega}{a}. Thus, our dynamical
calculations further support that the QAHC state at $\nu = 1/2$ represents a Chern insulator coexisting with
charge order in an enlarged unit cell.

\begin{figure*}[!htbp]
	\includegraphics[width=1\linewidth]{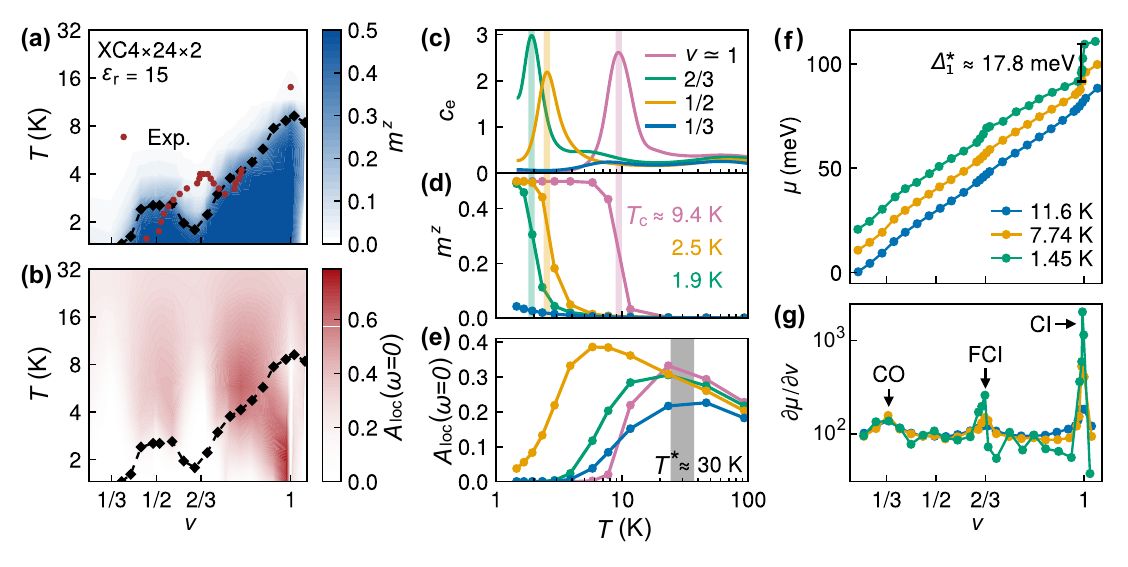}
	\caption{Finite-temperature properties for various fillings.
		(a) The magnetic moment $m^z$ and (b) the local density of states $A_{\rm loc}(\omega = 0)$ are displayed using contour plots. The experimentally determined FM transition temperatures, indicated by red dots in (a), are taken from Ref.~\cite{Cai2023Signatures}.	
		The black dashed line with diamonds shows the specific heat peaks at different fillings, which indicate the FM transition. Temperature evolution of (c) specific heat $c_e$, (d) magnet moment $m^z$, and (e) $A_{\rm loc}(\omega = 0)$ at typical fillings. (f) Chemical potential $\mu$ and (g) inverse compressibility 	 $\kappa^{-1}=\partial \mu/\partial \nu$ for various fillings $\nu$. The calculations are performed on a XC$4 \times 24\times 2$ lattice and with $\epsilon_\mathrm{r}=15$.
	}
	\label{Fig_FiniteT}
\end{figure*}

\section{Characteristic temperature scales}
In Fig.~\ref{Fig_FiniteT}, we present various finite-temperature properties calculated using the state-of-the-art thermal tensor network method, tanTRG~\cite{Li2023Tangent}. To characterize the moir\'e states, we compute the averaged magnetic moment $m^z = \frac{1}{2} \frac{\sum_{i}\expval{n_{i\uparrow}-n_{i\downarrow}}}{\sum_{i}\expval{n_{i\uparrow} + n_{i\downarrow}}}$ and the imaginary-time proxy  $A_{\rm loc}(\omega = 0) \approx \frac{2\beta}{N}\sum_{i,\sigma}\expval{c_{i\sigma}^\dagger(\beta/2)c_{i\sigma}}$, for the local density of states at the Fermi level~ \cite{Jiang2022NC, Samuel2017PANS, Li2023Tangent}. To reduce the boundary effects, we take averages over the bulk of the system. Contour plots of $m^z$ and $A_{\rm loc}(\omega = 0)$ versus temperature $T$ and filling $\nu$ are presented in Fig.~\ref{Fig_FiniteT}{a} and Fig.~\ref{Fig_FiniteT}{b}, respectively.

In Fig.~\ref{Fig_FiniteT}{c}, we show the specific heat results $c_e = \frac{1}{N}\left(\frac{\partial E}{\partial T}\right)_{\nu}$ for certain fillings $\nu$. Pronounced peaks observed at characteristic temperatures $T_c$ for $\nu = 1, 1/2,$ and $2/3$ signal the occurrence of phase transitions. Correspondingly, as depicted in Fig.~\ref{Fig_FiniteT}d, the magnetic moment \(m^z\) increases sharply near $T_c$ upon lowering the temperature and approaches the saturated value 0.5 for $T < T_c$, indicating a FM transition. In addition to the cases with $\nu = 1, 1/2, 2/3$, we also present results for the $\nu = 1/3$ case. Here, $m^z$ displays an increasing trend (though it has yet to converge to saturation), suggesting that the FM transition occurs below the lowest temperature of 1.45~K attained in the simulations.

As shown in Fig.~\ref{Fig_FiniteT}{a}, the Curie temperature of the FM transition is $T_c \gtrsim 2$~K for $0.4 \lesssim \nu \lesssim 1.05$, which closely aligns with the experimental observation that FM order can be observed at 1.6 K for $0.4 \lesssim \nu \lesssim 1.2$ \cite{Cai2023Signatures}. The FM transition temperatures exhibit an overall decreasing trend as $\nu$ deviates from 1, except at $\nu = 2/3$, where a dip is observed. This non-monotonic behavior is also qualitatively consistent with recent experimental findings~\cite{Cai2023Signatures}. Nevertheless, a quantitative discrepancy exists between the theoretical dip position at \(\nu \simeq 2/3\) in Fig.~\ref{Fig_FiniteT}{a} and the experimental observation, where the minimum is found at a filling factor of $\nu \approx 0.75$. This suggests that either certain interactions present in the experimental system is missing in our real-space model, or the model parameters --- particularly the $\epsilon_\mathrm{r}$, $\xi$, truncation length, or the system width $N_y$ --- need to be further optimized for a better agreement.

As temperature rises, even an insulating state can exhibit thermally activated charge excitations that enhance longitudinal conductivity, rendering a metal-like character to the system. This crossover in electrical transport behavior can be monitored through the local density of states at the Fermi level, $A_{\rm loc}(\omega=0)$. At low temperatures, $A_{\rm loc}(\omega = 0)$ is significantly suppressed for $\nu = 1$, 2/3, and 1/3, aligning with the insulating bulk states of the CI, FCI, and CO at these fillings, as have been demonstrated in DMRG calculations. A similar suppression is observed at $\nu = 1/2$, albeit at a lower temperature, possibly since the parameter $\epsilon_\mathrm{r} = 15$ is near the boundary between the QAHC and CO phases (see Fig.~\ref{Fig_nu12_D_3by4}b). On the other hand, in the small hole-doped regime to the left of $\nu = 1$ line, $A_{\rm loc}(\omega=0)$ clearly shows finite values till low temperature, which indicates the presence of metallic behaviors.

As shown in Fig.~\ref{Fig_FiniteT}{e}, the maximum of $A_{\rm loc}(\omega = 0)$ defines a characteristic thermal activation temperature $T^* \approx 30$ K for $\nu = 1, 2/3$ and $1/3$. Notably, the observed $T^* \simeq 30~\mathrm{K}$
is approximately 7 times lower than the charge gap $\Delta_1 \simeq 211.2~\mathrm{K}$ for $\nu = 1$ and about 3 times lower than $\Delta_{2/3} \simeq 97.4~\mathrm{K}$ for $\nu = 2/3$, highlighting a clear separation of energy scales. Our results may explain the experimental observations, where the thermal activation gaps extracted from transport measurements are also roughly 5 and 3 times smaller than the corresponding charge gaps obtained from local magnetometry probe in experiments~\cite{Redekop2024Direct, Park2023Observation}.

Moreover, in Fig.~\ref{Fig_FiniteT}{f}, we can estimate the charge gap from the calculated $\mu$-$\nu$ relation. A clear jump in the chemical potential appears for $\nu = 1$ at low temperatures, yielding a gap of $\Delta^*_1 \approx 17.8 $~meV. This value is in good agreement with the experimental value of $14 \pm 1$ meV~\cite{Redekop2024Direct}, as well as to the single-particle gap $\Delta_1 \approx 18.2$~meV estimated from our spectral function results in Fig.~\ref{Fig_GSAkomega}. The formation of a charge gap produces incompressible states, marked by peaks in the inverse compressibility, $\kappa^{-1} = \partial\mu/\partial\nu$. As shown in Fig.~\ref{Fig_FiniteT}g, $\kappa^{-1}$ derived from the $\mu$-$\nu$ relation exhibits pronounced peaks at $\nu = 1$, $2/3$, and $1/3$, corresponding to the CI, FCI, and CO ground states, respectively. Remarkably, these incompressible states have indeed been experimentally observed also by recognizing the $\kappa^{-1}$ peaks~\cite{Zeng2023Thermodynamic}.

Overall, we conclude that both energy scales, $\Delta_1^{(*)}$ and $T^*$, are significantly greater than the FM transition temperature. This suggests that the experimental observation of FCI and CI phases is ultimately limited by the FM transition temperature.

\section{Conclusion and Discussion}
In this study, we establish a moir\'e superlattice model and uniform framework to elucidate the intricate interplay between strong correlations and topology in tMoTe$_2$. Directly incorporating multiple bands and band-mixing effects, our accurate tensor-network calculations based on a realistic model successfully explain the existence of FCI states and account for related experimental observations, and predict the presence of QAHC states. Furthermore, our studies address a wide range of experimental observations, and shed light on both quantum dynamics and finite-temperature effects in the moir\'e system.

Our simulations align with recent experiments. Through a detailed exploration, we identify the FM$_z$ order in the range $0.2 \lesssim \nu \leq 1$, which covers the CI state at $\nu = 1$, and FCI states at fillings $\nu = 2/3$, 3/5, 4/7, 4/9, and 5/9 discovered in tMoTe$_2$~\cite{Zeng2023Thermodynamic, Cai2023Signatures, Xu2023Observation, Park2023Observation, Anderson2024Trion, Redekop2024Direct, Ji2024Local}. Our ground-state calculations unveil a rich $\epsilon_\mathrm{r}$-$\nu$ phase diagram, consisting of CI, FCI, QAHC, and CO phases. For instance, at $\nu = 1/3$ we observe a robust FCI phase for $15.5 < \epsilon_\mathrm{r} \leq 23$ and a GWC phase for $10 < \epsilon_\mathrm{r} \leq 15.5$, and the presence of the latter agrees with recent experimental results~\cite{Zeng2023Thermodynamic}. Our simulations suggest that, to observe the FCI and FM$_z$ phases at $\nu = 1/3$ in experiments,  it would require lower temperatures ($T < 1.45$ K) and a larger dielectric constant $\epsilon_\mathrm{r}$. Notably, nature of $\epsilon_\mathrm{r}$-driven quantum phase transitions between FCI and QAHC or CO phases have yet to be clarified and warrants future investigation. The QAHC states with integer Hall conductivity $\sigma_{xy} = 1$ at fractional fillings $\nu = 1/2$, 2/3, 3/5, 4/5, and nearby incommensurate fillings (e.g., $\nu\simeq 0.63$) also constitute a prediction remaining to be explored.

Our results indicate that the disruption of well-formed density structures destabilizes the QAHC phase, leading to its suppression or transformation into other competing states.
In experiments, the QAHC states have been identified in graphene-based systems with higher sample quality~\cite{Su2025moire, lu2024extended, aronson2024displacement, waters2024interplay, zhang2024commensurate}. Recent experiments on tMoTe$_2$ have revealed substantial real-space disorder through direct magnetic imaging~\cite{Redekop2024Direct}, highlighting the need for high-quality samples to realize QAHCs in this system. Moreover, we conjecture that some of the experimentally observed FCI states at fillings such as $\nu = 1/2$, 2/3, and 3/5 may originate from the ``melting'' of CO in the QAHC states --- thermal fluctuations may drive a transition from QAHC to FCI states~\cite{lu2024extended}. Therefore, constructing the complete finite-temperature phase diagram for model~(Eq.~(\ref{Eq_Model})) would not only clarify the nature of the topological Hall states in tMoTe$_2$ but also offer valuable guidance for future experimental work.

The influence of external fields on QAHCs also calls for further studies. QAHCs can be tuned into FCI states via applied current~\cite{lu2024extended}, perpendicular electric fields (displacement fields)~\cite{lu2024extended, aronson2024displacement, waters2024interplay, zhang2024commensurate}, and other experimental parameters. In particular, the displacement field, which introduces an onsite potential difference $u_{\rm D}$ between the two sublattices, is a commonly used tunable parameter
in experiments and can drive a quantum phase transition from topological Hall states to trivial insulators in tMoTe$_2$~\cite{Cai2023Signatures, Zeng2023Thermodynamic, Xu2023Observation, Park2023Observation}. Thus, the effects of the displacement field on various states, including the CI at $\nu = 1$, FCI at $\nu = 2/3$~\cite{sharma2024topological}, and QAHC at $\nu = 1/2$, remain to be explored. The insights gained from real-space models Eq.~(\ref{Eq_Model}) may offer valuable guidance also for studying other moir\'e systems with similar honeycomb superlattices~\cite{Andrei2020,Mak2022Semiconductor},
such as twisted WSe$_2$, AB-stacked MoTe$_2$/WSe$_2$, and AA-stacked bilayer graphene, etc. In addition, an external magnetic field can effectively suppress spin fluctuations, potentially influence both the ground-state and finite-temperature properties~\cite{park2025observation}. Furthermore, a broader range of experimentally tunable parameters~\cite{Mak2022Semiconductor} --- such as twist angle, filling, dielectric constant, mechanical strain, optical excitation, and screening length --- also remain to be investigated using the real-space approach introduced in the present study.

Beyond the ground-state property, our many-body results reveal three characteristic energy scales for topological Hall states: the FM transition temperature ($T_c$), charge activation temperature ($T^*$), and the zero-temperature charge gap ($\Delta_{\nu}$).  The observed charge gaps at low temperature have been found to be about 7 (3) times greater than the thermal activation temperatures for $\nu=1$ ($\nu=2/3$), which is consistent with experiments and resolves a key experimental contradiction \cite{Redekop2024Direct, Park2023Observation}. One possible explanation for this observation attributes it to the formation of in-gap excitonic bound states. Their proliferation at finite temperature could induce strong fluctuations and renormalization effects, lowering the thermal activation threshold for excitations. Similar phenomena have been observed numerically in the topological Mott insulator~\cite{Chen2021Realization, Lin2022Exciton, Lu2024Thermodynamic}  and bosonic FCI phases~\cite{lu2024vestigial}. Further investigation into the collective excitations in the topological Hall states --- such as excitons~\cite{Mak2022Semiconductor},
magneto-rotons~\cite{Lu_2024}, possible graviton modes~\cite{Liang2024Evidence}, and fractional excitations~\cite{Nakamura2020Direct} --- and the comparisons with spectroscopic measurements are left for future studies.
Finally, the methodology presented here can, in principle, be used to explore exotic non-Abelian fractional excitations --- particularly in the recently discovered fractional quantum spin Hall state~\cite{Kang2024Evidence}.

\textit{Note added.---} During preparing our manuscript, we note an experimental work~\cite{xu2025signature} that observed a compressible metallic region between $\nu=2/3$ to 1 that breaks spin and valley symmetry, as well as integer quantum anomalous Hall crystals near $\nu=2/3$ and $3/5$ fillings. These experimental observations are consistent with our theoretical results here. The unconventional superconductivity observed in this system --- proposed to be chiral superconductivity through weak-coupling approach on a continuum model~\cite{xu2025chiralsc} --- is now ready to be investigated using precision many-body calculations on the moir\'e superlattice model established here.

\section*{Conflict of interest}
The authors declare that they have no conflict of interest.

\section*{Acknowledgments} Jialin Chen, Qiaoyi Li, and Wei Li are indebted to Bin-Bin Chen and Hongyu Lu for stimulating discussions. The iDMRG calculation is based on TeNPy \cite{tenpy2024}. The dynamical and finite-temperature calculations are based on TensorKit.jl~\cite{Devos2025TensorKit} and  FiniteMPS.jl~(https://github.com/Qiaoyi-Li/FiniteMPS.jl). This work was supported by the National Natural Science Foundation of China (12222412, 12047503), National Key Projects for Research and Development of China (2024YFA1 409200), Innovation Program for Quantum Science and Technology (2021ZD0301900), Strategic Priority Research Program of Chinese Academy of Sciences (XDB1270100), and the Postdoctoral Fellowship Program of CPSF (GZB202 40772). We thank the HPC-ITP for the technical support and generous allocation of CPU time. Xiaoyu Wang acknowledges financial support from the National High Magnetic Field Laboratory through National Science Foundation (NSF) Grant No. DMR-2128556, and the State of Florida.

\section*{Author contributions}
Wei Li and Xiaoyu Wang initiated this work. Xiaoyu Wang constructed the lattice  model. Wei Li, Jialin Chen, and Qiaoyi Li conducted the model analysis and tensor-network calculations. All authors contributed to the analysis of the results and the preparation of the draft.

\bibliography{REFbib.bib}


\newpage
\clearpage
\onecolumngrid
\mbox{}
\begin{center}
	
	{\large Supplementary Materials for} $\,$ \\
	\bigskip
	\textbf{\large{Fractional Chern Insulator and Quantum Anomalous Hall Crystal in Twisted MoTe$_2$}} \\
	Chen \textit{et al}.
\end{center}

\date{\today}

\setcounter{section}{0}
\setcounter{figure}{0}
\setcounter{equation}{0}
\renewcommand{\theequation}{S\arabic{equation}}
\renewcommand{\thefigure}{S\arabic{figure}}
\setcounter{secnumdepth}{3}

\section{Model Hamiltonian}{\label{App_Models}}
Here we present a derivation of the model Hamiltonian of tMoTe$_2$ from the continuum electronic model. The experimentally relevant twist angle for this work is in the range of $3.4^\circ\sim 3.9^\circ$, where fractionalized quantum anomalous Hall effect has been observed \cite{Cai2023Signatures,Zeng2023Thermodynamic,Xu2023Observation,Park2023Observation,Ji2024Local,Anderson2024Trion, Redekop2024Direct}. 

\subsection{Continuum model of twisted bilayer MoTe$_2$}

Monolayer MoTe$_2$ features strong spin-orbit coupling \cite{DiXiao2012}, such that the spin (${\uparrow,\downarrow }$) and valley (${\bK,\bK'}$) quantum numbers are locked. We hereby use the convention of $\{\bK,\uparrow\}$ and $\{\bK',\downarrow\}$ to label the electronic states.

\begin{figure*}[h]
	\centering
	\includegraphics[width=0.9\linewidth]{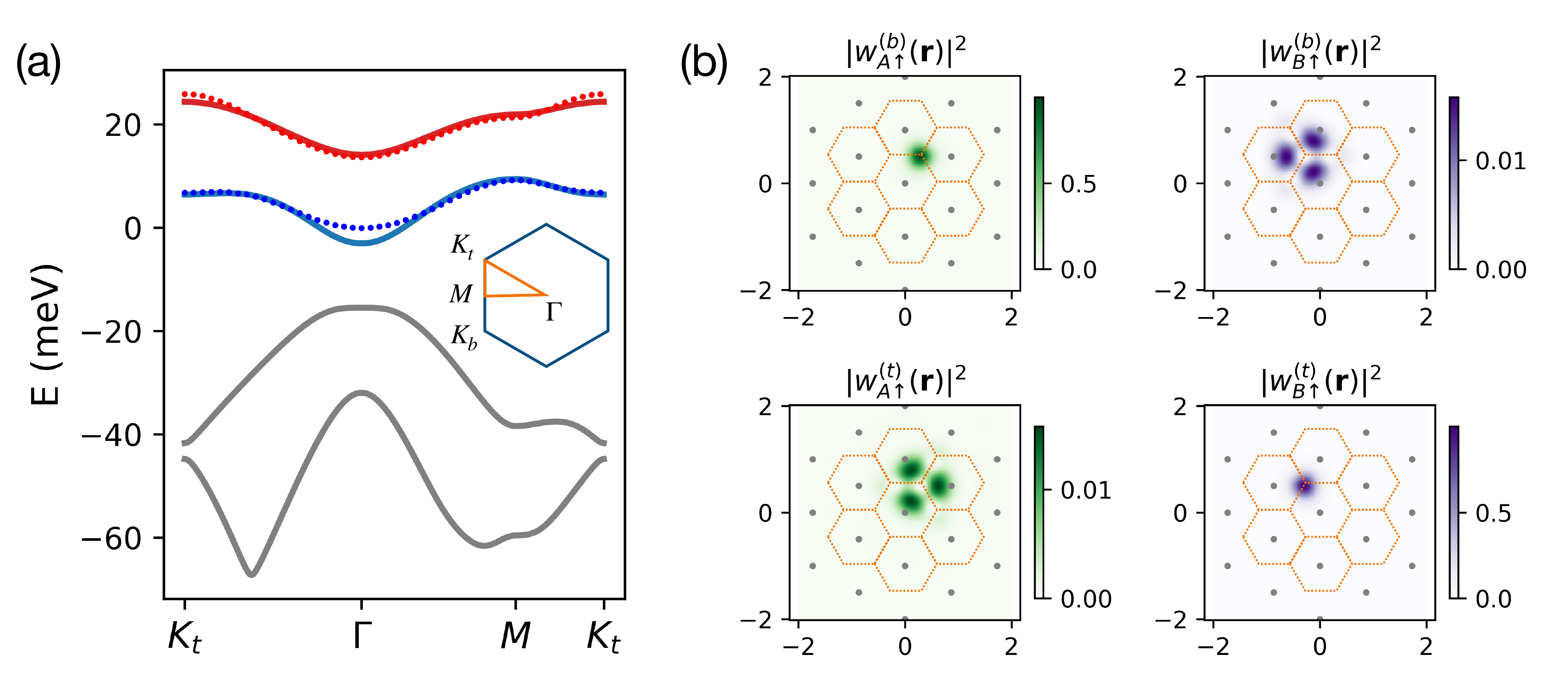}
	\caption{(a) Solid lines: electronic dispersions of valley $\bK$ and spin $\uparrow$ in the moir\'e Brillouin zone. The two colored bands 
		(blue and red), along with their time reversed partners in valley $\bK'$ and spin $\downarrow$, are the most relevant low energy bands 
		for experiments. The inset depicts the moir\'e Brillouin zone and the linecut (orange) along which the dispersion is drawn. Dotted lines: 
		tight-binding fits to the dispersions of the upper two bands. (b) Real space electronic density profile of the pair of exponentially localized 
		Wannier orbitals, normalized with respect to the peak density. The real space coordinates are measured with the length of the moir\'e unit 
		cell vector set to 1. The Wannier orbitals reside mostly at the honeycomb lattice sites, corresponding to the XM/MX stacking regions of the 
		tMoTe$_2$. The pair of Wannier orbitals, labeled by $A$ and $B$ respectively, are also predominantly layer resolved. 
		Specifically, $\psi_{A\uparrow}$ reside predominantly on the bottom layer (upper left panel), whereas $\psi_{B\uparrow}$ resides 
		predominantly on the top layer (lower right panel).}
	\label{fig:figS1}
\end{figure*}

Focusing on $\{\bK,\uparrow\}$, the relevant low energy degrees of freedom are the valence bands of the two MoTe$_2$ layers. The moir\'e superlattice potential hybridizes the two layers, with a minimal model given by the following continuum Hamiltonian \cite{Wu2019Topological}: 
\begin{equation} \label{eq:cont_nonint}
	\begin{split}
		{H}_{\uparrow} = \int \mathrm{d}^2 \br
		\begin{pmatrix}
			\psi^{\dagger}_{b \uparrow}(\br) & \psi^\dagger_{t \uparrow}(\br)
		\end{pmatrix}  
		\begin{pmatrix}
			{h}_{b}({\bp}) + \Delta_{b}(\br) + \frac{u_D}{2} & \Delta_{T}(\br) \\
			\Delta_{T}^{*}(\br) & {h}_{t}({\bp}) + \Delta_{t}(\br) - \frac{u_D}{2}
		\end{pmatrix}
		\begin{pmatrix}
			\psi_{b \uparrow}(\br)\\ \psi_{t \uparrow}(\br)
		\end{pmatrix}.
	\end{split}
\end{equation}
Here $\psi^{\dagger}_{l \uparrow}(\br)$ creates an electron in layer $l=b, t$ , with spin $\uparrow$ and in valley $\bK$, and at position $\br$. For convenience we drop the valley subscript and label the electrons by their spin quantum number. ${h}_{l}({\bp}) = -({\mathbf{p}}- \hbar \bK_{l})^2/2m^*$ is kinetic energy, with ${\bp}=-i\hbar \nabla_{\br}$ and effective mass $m^{*}$. $\bK_{l}$ points sit at the corners of the hexagonal moir\'e Brillouin zone (see Fig.~\ref{fig:figS1}(a) inset). We describe the effect of displacement field $\mathbf{D}$ by adding a potential difference $u_D$ between two layers. $\Delta_{b(t)}(\br) = 2 V \sum_{j = 1, 3, 5} \cos(\bG_j \cdot \br \pm \phi)$ is the intralayer potential for bottom (top) layer, and $\Delta_T(\br) = w(1 + e^{-i \bG_2 \cdot \br} + e^{-i \bG_3 \cdot \br})$ is interlayer tunneling potential. $\bG_{i=1,\dots 6}$ are the six smallest reciprocal lattice vectors, with $\bG_1=\frac{4\pi}{\sqrt{3}L_M}\hat{x}$ along the $+x$ direction, and $\bG_{i+1}$ is the vector obtained by rotating $\bG_{i}$ $60^\circ$ counterclockwise. Here $L_M\approx a_0/\theta$ is the moir\'e unit vector length, with $\theta$ the twist angle and $a_0$ the MoTe$_2$ lattice spacing. 

\begin{table}[h]
	\centering
	\begin{tabular}{l|c|c|c|c|c}
		\hline
		& $a_0$  [\AA] & $m^*$ [$m_e$] & $V$ [meV] & $\phi$ & $\omega$ [meV] \\ \hline 
		Ref.~\cite{Wu2019Topological} & 3.47 & 0.62 & 8.0 & -$89.6^\circ$ & -8.5 \\ \hline 
		Ref.~\cite{wang2023topology} & 3.55 & 0.62 & 17.0 & $107.7^\circ$ & -16.0 \\  \hline 
		Ref.~\cite{Wang2024Fractional} & 3.52 & 0.6 & 20.8 & $107.7^\circ$ & -23.8 \\ \hline 
		Ref.~\cite{Reddy2023Fractional} & 3.52 & 0.62 & 7.5 & -$100^\circ$ & -11.3 \\ \hline   
	\end{tabular}
	\caption{Continuum model parameters from literature.}
	\label{tab:model_params}
\end{table}

The continuum model parameters are obtained by fitting the continuum model dispersion to DFT calculations 
\cite{Wu2019Topological, Reddy2023Fractional,wang2023topology,Wang2024Fractional,WuQuanSheng2024}, 
with different sets of parameters obtained depending on the literature (see Table~\ref{tab:model_params}).
Besides, Ref.~\cite{WuQuanSheng2024} has included terms beyond those included in Eq.~(\ref{eq:cont_nonint}). 
For concreteness, we utilize the parameters specified in Ref.~\cite{wang2023topology} and a twist angle of 
$\theta=3.89^\circ$. Although our tensor-network results are expected to be qualitatively consistent with other model 
parameters, a likely exception is Ref.\cite{Reddy2023Fractional}, as it predicts a different sequence of moir\'e 
band Chern numbers compared to the other DFT calculations. Figure~\ref{fig:figS1}(a) shows the energy 
dispersions in valley $\bK$ and spin $\uparrow$ in the moir\'e Brillouin zone. Energy dispersions in valley 
$\bK'$ and spin $\downarrow$ are related by time reversal symmetry. Experimentally, FCI states are achieved 
by doping holes into the valence bands, corresponding to partial fillings of the band colored in red.

\subsection{Tight binding fit to the continuum model}
With the exception of Ref.~\cite{Reddy2023Fractional}, other DFT calculations predict the upper two moir\'e bands of a given valley carry Chern numbers $+1$ and $-1$ respectively, including the one we used for this work. This allows us to construct exponentially localized Wannier orbitals without suffering from topological obstruction of a band composite with a finite Chern number \cite{Soluyanov2011Wannier}. Using the trial state procedure discussed in Ref.~\cite{Soluyanov2011Wannier}, we construct exponentially localized Wannier orbitals to describe the Hilbert space spanned by the upper two valence bands per valley, see Fig.~\ref{fig:figS1}(b). The two Wannier orbitals per valley reside predominantly on the honeycomb lattice, which is the dual lattice of the moir\'e triangular lattice made of AA stackings. We hereby label these states as $w_{A}(\br)$ and $w_{B}(\br)$ respectively, with the subscripts $A,B$ denoting the center of the Wannier orbitals. Separately, these two orbitals are layer polarized, with $w_{A(B)}(\br)$ having most orbital weights on the bottom (top) layer. 

In summary, the two Chern bands per valley can be represented using the Wannier orbital basis states, $\{w_{i,\sigma}(\br)\}$, where $\sigma=\uparrow,\downarrow$. $i$ denotes the Wannier orbital center at $\bR+\tau$, where $\tau=A,B$ is the position of a sublattice inside a moir\'e unit cell centered at $\bR=m\bL_1+n\bL_2$. $\bL_{i=1,2}$ are moir\'e unit cell vectors. Note that the Wannier orbital is a spinor in the layer basis. 

\begin{figure*}[!htbp]
	\centering
	\includegraphics[width=\linewidth]{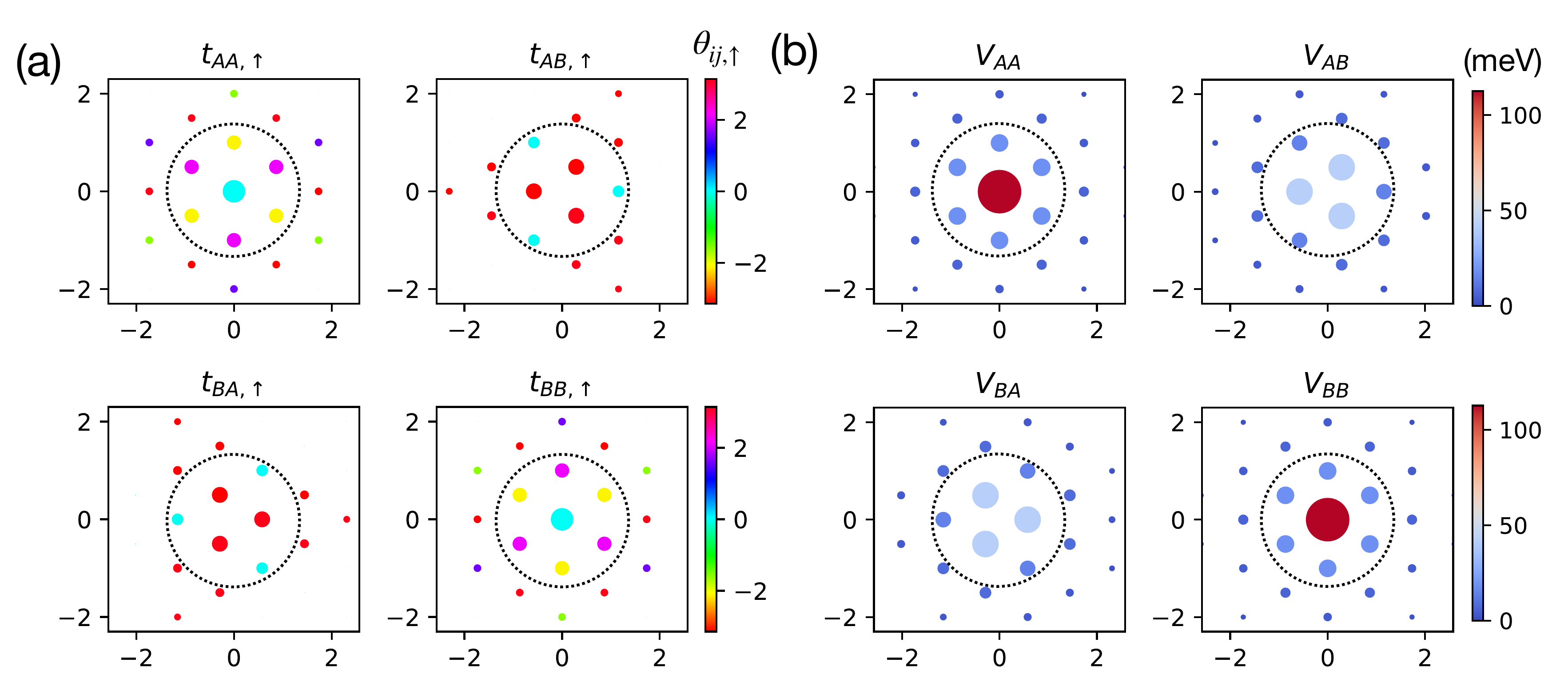}
	\caption{(a) Hopping amplitudes $t_{ij,\sigma}$ in real space, where the length of the moir\'e unit cell vector is set to 1. 
		Here we use the area size of the dots to denote the magnitude and the color to denote the phase (in unit of radian), such that $t_{ij,\sigma}
		\equiv |t_{ij,\sigma}|e^{i\theta_{ij,\sigma}}$. The onsite terms $\mu_{A,B}$ are depicted as cyan dots at $(0,0)$ in the upper 
		left and lower right panels. In  tensor-network simulations we kept hopping terms enclosed by the black dashed circles. (b) Projected 
		density-density interactions. The strength of the interactions are denoted by both dot size and color. The interaction strengths 
		are obtained for screening distance $\xi=30$ nm and relative dielectric constant $\epsilon_r=10$. In  tensor-network simulations we 
		keep interaction terms enclosed by the black dashed circles (and with a shorter screening length $\xi$).}
	\label{fig:figs2}
\end{figure*}

The tight-binding model is therefore given as: 
\begin{equation}{\label{SM_Eq_TB}}
	H_{\rm TB} = \sum_{ij,\sigma} t_{ij,\sigma} c^{\dagger}_{i\sigma}c_{j\sigma} - \sum_{i,\sigma} \mu_{i\sigma} c^{\dagger}_{i\sigma}c_{i\sigma},
\end{equation}
where $c^{\dagger}_{i\sigma}$ creates an electron in the Wannier orbital labeled by $i$ and spin $\sigma$. Due to time reversal symmetry $t_{ij,\downarrow}=t_{ij,\uparrow}^*$. The hopping amplitudes $t_{ij,\uparrow}$ and the onsite terms $\mu_i$ are shown in Fig.~\ref{fig:figs2}(a).

In this work, we truncate the hopping terms to the third nearest neighbor, corresponding to the the black dashed circle in Fig.~\ref{fig:figs2}(a). 
The tight-binding dispersion is plotted as dotted line in Fig.~\ref{fig:figS1}(a), showing its comparison with the continuum model.

\subsection{Screened Coulomb interaction}
We consider a dual-gate screened Coulomb interaction given by 
\begin{equation} \label{eq:contCoulomb}
	H_U = \frac{1}{2A} \iint \mathrm{d}^2\br \mathrm{d}^2 \br' V(\br-\br') \hat{P}\rho(\br)\hat{P} \hat{P}\rho(\br')\hat{P},
\end{equation}
where $A$ is the area of the system. The interaction coefficient and its Fourier transform are given by 
\begin{equation}
	\notag
	\begin{split}
		& V(\br-\br') = \frac{e^2}{4\pi \epsilon_0 \epsilon_r |\br-\br'|} e^{-|\br-\br'|/\xi}, \\
		&\ V_{\bq} = \frac{1}{A}\frac{e^2}{2\epsilon_0\epsilon_r}\frac{\tanh(|\bq|\xi)}{|\bq|},
	\end{split}
\end{equation}
where $\epsilon_0$ is the vacuum permittivity, $\epsilon_r$ is the relative dielectric constant, 
$\xi$ is the screening length. $\rho(\br)$ is the electron density operator:
\begin{equation}
	\notag
	\rho(\br) = \sum_{l,\sigma}\psi^{\dagger}_{l,\sigma}(\br)\psi_{l,\sigma}(\br). 
\end{equation}
$\hat{P}$ is the projector onto the narrow bands of interest, i.e., the colored bands in Fig.~\ref{fig:figS1}(a). 
In the Wannier orbital basis, the projected density operator becomes: 
\begin{equation}
	\notag
	\hat{P} \rho(\br) \hat{P} = \sum_{ij,\sigma} \left[ w^\dagger_{i,\sigma}(\br) w_{j,\sigma}(\br)\right] c^{\dagger}_{i\sigma}c_{j\sigma}.
\end{equation}

Finally, the Coulomb interaction projected onto the low energy Hilbert space is given as: 
\begin{equation} \label{eq:CoulombProjection}
	H_U = \frac{1}{2}\sum_{ijkl,\sigma\sigma'} V_{ij,kl}^{\sigma,\sigma'} c^{\dagger}_{i\sigma}c_{j\sigma}c^{\dagger}_{k\sigma'}c_{l\sigma'},
\end{equation}
where
\begin{equation}
	\notag
	\begin{split}
		V_{ij,kl}^{\sigma,\sigma'} = \frac{1}{A}\iint \mathrm{d}^2 \br \mathrm{d}^2 \br' V(\br-\br')  \left[ w^\dagger_{i,\sigma}(\br) w_{j,\sigma}(\br)\right] \left[ w^\dagger_{k,\sigma'}(\br') w_{l,\sigma'}(\br')\right].
	\end{split}
\end{equation}
Evaluating above integration directly in real space can be challenging. In practice, we switch to Fourier space and calculate the following: 
\begin{equation}
	\notag
	\begin{split}
		V_{ij,kl}^{\sigma,\sigma'} = \sum_{\bq} V_{\bq} \left[\int \mathrm{d}^2 \br w^\dagger_{i,\sigma}(\br) e^{-i\bq\cdot\br} w_{j,\sigma}(\br)\right] \left[ \int \mathrm{d}^2 \br' w^\dagger_{k,\sigma'}(\br') e^{i\bq\cdot\br'} w_{l,\sigma'}(\br')\right].
	\end{split}
\end{equation}
The two bracketed integrals can be evaluated conveniently by representing Wannier orbitals using the Bloch basis states.

Given the exponential localization of the Wannier orbitals, interaction coefficients with $i\neq j$ or $k\neq l$ (assisted hopping terms) 
are small compared to density-density interactions with $i=j$ and $k=l$. For relative dielectric constant $\epsilon_r=10$ and screening 
distance $\xi \approx 5.77 L_m \simeq 30$~nm, interaction coefficients with $i\neq j$ or $k\neq l$ are all $\le 1~{\rm meV}$. We hereby 
drop these terms in our simulations, and approximate the projected Coulomb interaction by density-density interactions,
\begin{equation}{\label{SM_Eq_Coulomn_terms}}
	\begin{split}
		H_U \approx \frac{1}{2}\sum_{ij,\sigma\sigma'} V_{ii,jj}^{\sigma,\sigma'} n_{i\sigma}n_{j\sigma'} 
		\equiv U\sum_{i\sigma}n_{i\uparrow}n_{i\downarrow} + \sum_{i<j,\sigma\sigma'} V_{ij} n_{i \sigma} n_{j\sigma'}.
	\end{split}
\end{equation}

Fig.~\ref{fig:figs2}(b) shows the interaction coefficients $\{U,V_{ij}\}$. Due to the exponential localization of the Wannier orbitals, 
to a good degree of quantitative accuracy, we find the following approximate expression for the density density interactions: 
\begin{equation}
	\notag
	\begin{split}
		U &\approx \frac{1192.71}{\epsilon_r} {\rm meV},\\  V(\br) & \approx \frac{276.92}{\epsilon_r} \frac{L_m}{|\br|} \exp 
		\left( -\frac{|\br|/L_m}{\xi/L_m}\right){\rm meV},
	\end{split}
\end{equation}
where $L_m$ is the length of the moir\'e unit cell vector. This analytic expression is used in the  tensor-network simulations. Furthermore, 
due to system size constraints, we cut off the interaction terms beyond the dotted circles shown in Fig.~\ref{fig:figs2}(b). In the 
main text, we choose a shorter screening length of $\xi = 2 L_m$, which results in an even smaller truncation error for the long-range 
couplings.

In the main text, we truncate both the hopping and interaction terms up to the third nearest neighbor. Since the truncated hopping terms are small (e.g., \( t_4 = 0.319 \) meV and \( t_5 = 0.180 \) meV), the truncation error primarily arises from the interaction terms. To assess the effects of truncation, we calculate the FM\(_z\) gaps for models with \( \epsilon_r = 20 \), \( \nu = 1 \), first-three-nearest-neighbor hopping terms, and interaction terms truncated at the 2nd, 3rd, and 5th nearest neighbors. Using iDMRG with a \( 2 \times 4 \times 4 \) unit cell, we obtain FM\(_z\) gaps per unit cell of \( \Delta E^{\text{2nd}} \simeq 1.187 \) meV, \( \Delta E^{\text{3rd}} \simeq 1.218 \) meV, and \( \Delta E^{\text{5th}} \simeq 1.233 \) meV, demonstrating good consistency. These results indicate that the gap increases slightly when longer-range terms are included, confirming that truncation at the third nearest neighbor already provides a good approximation---hence its adoption in the main text.  

It is important to note that truncation must be performed at a point where inter- and intra-sublattice interaction terms remain balanced; otherwise, it may lead to a sublattice-polarized phase. For example, since \( V_1 \), \( V_3 \), and \( V_4 \) are inter-sublattice interactions while \( V_2 \) and \( V_5 \) are intra-sublattice interactions, retaining only \( V_1 \) (or terms up to \( V_4 \)) would result in a sublattice-polarized antiferromagnetic ground state—an unphysical outcome that should be avoided.

\begin{figure*}[!hbtp]
	\includegraphics[width=1.0\linewidth]{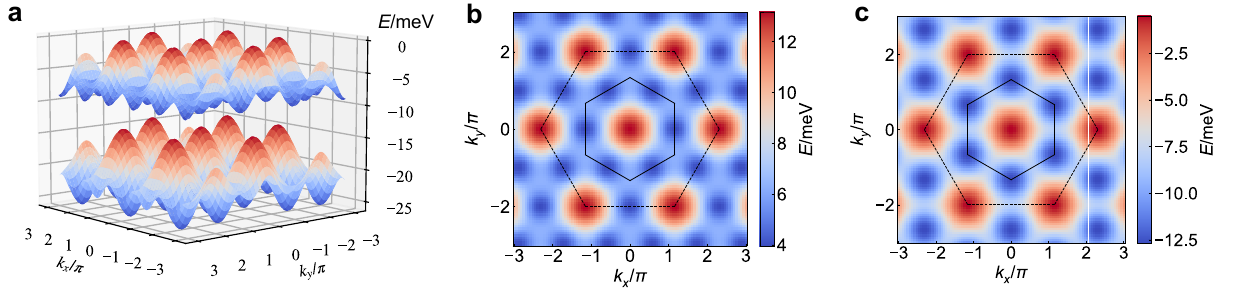} 
	\caption{\textbf{a} Band structure of the tight-binding model~(\ref{SM_Eq_TB}) with hopping terms up to the third nearest neighbor 
		after particle-hole transformation. Panels \textbf{b} and \textbf{c} represent the upper and lower bands, respectively. The smaller 
		(larger) honeycomb with black solid (dashed) line represents the first (enlarged) Brillouin zone. 
	}
	\label{SM_Fig_Band_t3V3}
\end{figure*}

\subsection{Particle-hole transformation}
The parameters in Eq.~(\ref{eq:cont_nonint}) are assumed to lead to experimentally measured band dispersions at the reference state 
$\ket{\Omega}$ corresponding to hole filling 0, i.e., when the chemical potential is placed above the moir\'e bands depicted in 
Fig.~\ref{fig:figS1}(a). However, with respect to $\ket{\Omega}$, the projected Coulomb interaction in Eq.~(\ref{eq:CoulombProjection}) 
will renormalize the band dispersions away from that obtained using Eq.~(\ref{eq:cont_nonint}). This leads to double counting, 
and a quadratic order counter term needs to be added to avert this effect. 

Alternatively, we can perform a particle-hole (PH) transformation $d^\dagger_{i\sigma} = c_{i\sigma}$ which inverts the non-interacting 
band structure (as well as the hopping parameters and chemical potential), while simultaneously we preserve the same functional 
form for the projected density density interactions but with $n_{i\sigma}=d^{\dagger}_{i\sigma}d_{i\sigma}$. Such a transformation 
has been used in previous theoretical studies on tMoTe$_2$ (see, e.g., Ref.~\cite{Reddy2023Fractional}). Here $\ket{\Omega}$ will 
be a vacuum state for $d$ fermions, and therefore the density-density interaction will {\it not} lead to band renormalizations. 

Therefore, instead of stating our results for hole fillings $-1\le\nu_h<0$, we report results for electron fillings $0<\nu\le 1$. 
Accordingly, we present the band structure of the model described by Eq.~(\ref{SM_Eq_TB}) in Fig.~\ref{SM_Fig_Band_t3V3}, 
which includes hopping terms up to the third nearest neighbors following the PH transformation. Without interactions, 
electrons only fill the lower band [see, Fig.~\ref{SM_Fig_Band_t3V3}(c)] when filling factor $\nu\le 1$.

It should be noted that the PH transformation employed in our work serves as a strict mathematical tool for reformulating the Hamiltonian, and it is conceptually distinct from any approximate PH symmetry in the flat bands. Since our narrow-band model inherently lacks PH symmetry, it cannot be mapped to the lowest Landau level. This intrinsic asymmetry naturally explains the strong PH asymmetry observed in experiments. Consistently, our DMRG results display the same trend: as shown in the phase diagrams of Figs.~1 and~6 in the main text and Fig.~\ref{Other_FCI}, topological phases (CI, FCI, and QAHC) and FM$_z$ states are more robust for $\nu > 1/2$ than for their PH counterparts at $1-\nu$. For example, the $\epsilon_r$ stability ranges at $\nu = 2/3$ and $5/9$ are broader than those at $\nu = 1/3$ and $4/9$, respectively. Moreover, robust topological phases are observed at $\nu = 4/7$, $4/5$, and $3/5$, but not at the corresponding fillings $\nu = 3/7$, $1/5$, and $2/5$.

\section{Details of Methods}
\subsection{Model simplification.}
Since the onsite Coulomb interaction $U$ exceeds the hopping parameters $t_1, t_2, t_3$ by over
an order of magnitude, we can project out double occupancy at low temperatures. The numerical 
benchmarks between two models are shown in section~\ref{SM_FiniteT}. Then we arrive 
at a spinful $t$-$V$ model whose Hamiltonian reads  
\begin{equation}{\label{Eq_Model_tJ}}
	\begin{split}
		H = & \sum_{i<j} \big( t_{ij } P_i c^\dagger_{i\uparrow} c_{j\uparrow} P_j 
		+ t^*_{ij} P_i c^\dagger_{i\downarrow} c_{j,\downarrow} P_j + h.c. \big)  
		-\mu\sum_{i}  n_{i} P_i +  \sum_{  i<j } V_{ij}  P_i n_i  n_j P_j,
	\end{split}
\end{equation}
where $P_i= 1 - n_{i\uparrow} n_{i\downarrow}$ is a projector on site $i$. The spin-spin coupling $J$ can 
be neglected as $U \gg |t_{ij}|$ and thus {$|t_{ij}|^2/U \ll 1$}~\cite{Cheng2024Maximally}. This projected 
spinful $t$-$V$ model is utilized in our finite-temperature (and also part of ground-state) calculations.

In the calculations of fully spin-polarized cases, e.g., FCI and QAHC states, we further simplify the model 
to a spinless version. This is justified as Eq.~(\ref{Eq_Model_tJ}) conserves electron numbers for both spin 
flavors, and the spinless Hamiltonian is given by 
\begin{equation}{\label{Eq_Model_spinless}}
	\begin{split}
		H = & \sum_{i<j} \big( t_{ij} c^\dagger_{i} c_{j} + h.c. \big)  
		-\mu  \sum_{i}  n_{i}  + \sum_{i<j} V_{ij} n_{i} n_{j},
	\end{split}
\end{equation}
where $c^\dagger_i$ creates an electron at site $i$, and $n_i = c^\dagger_i c_i$ is the particle number 
operator. This spinless $t$-$V$ model is employed in the calculation of Hall conductivity, and its validity 
is confirmed by comparing the results with those from the Hubbard-type model~(Eq.~(1) in the main text) or the 
spinful $t$-$V$ model~(\ref{Eq_Model_tJ}).

The three models and their usage in various simulations are summarized in Table~\ref{tab:model_for_TNW}.   
\begin{table}[h]
	\centering
	\begin{tabular}{|c|c|c|c|c|c|}
		\hline
		Model & No. & $d$ & $T=0$ & $T>0$ & Dynamics \\ \hline 
		Hubbard-type & Eq.~(1) & 4  & \checkmark & \checkmark & $\times$ \\ \hline 
		spinful $t$-$V$ & Eq.~(\ref{Eq_Model_tJ}) &  3 & \checkmark & \checkmark & $\times$ \\ \hline 
		spinless $t$-$V$  & Eq.~(\ref{Eq_Model_spinless}) &   2 & \checkmark & $\times$ & \checkmark \\ \hline 
	\end{tabular}
	\caption{Real-space models used in tensor-network calculations. Here $d$ refers to the dimension of local Hilbert space. }
	\label{tab:model_for_TNW}
\end{table}

\subsection{Density matrix renormalization group.}
For the ground state, we utilize both finite and infinite density matrix renormalization group (DMRG and iDMRG)
\cite{White1992Density,mcculloch2008infinite,SCHOLLWOCK2011The,tenpy2024}.
The iDMRG method is based on the infinite matrix product states (MPS) constructed from an enlarged unit cell with size $N_y \times N^c_x
\times 2$. This approach allows us to explore the periodicity of the state by varying the size of unit cell, with the 
consistency of results confirmed through comparison to finite DMRG results. In practice, we retain a bond 
dimension up to $\chi=4,096$ in iDMRG simulations, and the truncation errors $\lesssim 2\times 10^{-5}$, 
thereby guaranteeing very good convergence and thus high accuracy of the results. In finite-size DMRG calculations, we fine-tune the total particle number to maintain bulk filling at $\nu \simeq p/q$ for integer 
$p$ and $q$.

\subsection{Time-dependent variational principle.}
After obtaining the ground state, we employ real-time 
evolution with the TDVP approach~\cite{Haegeman2011PRLTime,Haegeman2016PRBUnifying} to compute 
the retarded Green's function and subsequently derive the single-particle spectral function via Fourier 
transformation. In the computation of dynamical properties, it is necessary to determine the chemical 
potential $\mu$ beforehand in order to fix the position of the Fermi level $(\omega = 0)$. In the present 
study, we estimate $\mu$ based on low-temperature calculations (practically at $T \simeq 1.45$K), utilizing 
the $\mu$-$\nu$ relation (see, e.g., Fig.~6\textbf{f} in the main text). In practice, a Parzen window is applied 
to suppress non-physical oscillations~\cite{Kuehner1999PRBDynamical,Li2022PRRDetecting}, achieving 
an energy resolution of $2.67$~meV (with evolution time up to 3~meV$^{-1}$). We choose time step $\delta 
t =1/16$~meV$^{-1}$ and bond dimension $\chi = 1000$, which renders accurate results in the TDVP 
calculations. The data convergence and more technical details are presented in the section~\ref{SM_Sec_Spectral}. 

\subsection{Tangent-space tensor renormalization group.} 
The tanTRG method~\cite{Li2023Tangent} 
is employed to study the finite-temperature properties of the system. An accurate matrix product operator 
representation of the density matrix, with bond dimension $\chi$ up to 2000, enables us to investigate systems 
with sizes up to XC$4\times 24\times2$ and down to a low temperature of $T \simeq 1.45$~K (i.e., $T/t_1
\simeq 0.0375$). In our tanTRG calculations, we employ the grand canonical ensemble and implement a 
specially designed chemical potential $\mu$-tuning technique to precisely control the particle number. To 
study the FM transition, we introduce a small pinning field $h = 0.01$ to stabilize the FM order in 
the finite-size systems. To improve the efficiency, we implement Abelian spin and/or charge symmetries in 
the simulations and adopt controlled bond expansion techniques~\cite{Gleis2023PRLControlled,Li2024PRLTime} 
to speed up the finite-temperature calculations.

\section{Supplementary ground-state results}

\subsection{Magnetic gaps for various fillings}
\begin{figure}[!h]
	\includegraphics[width=0.52\linewidth]{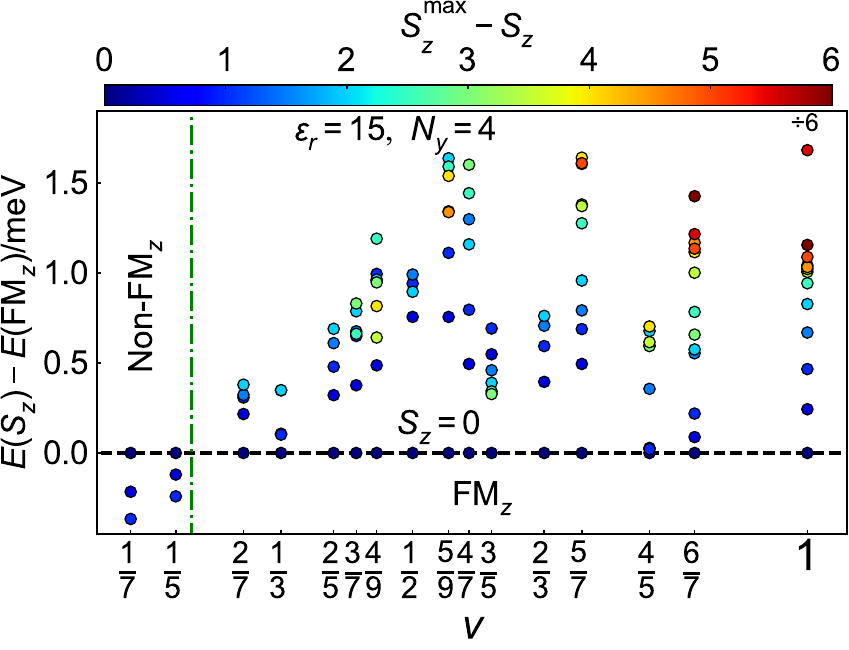} 
	\caption{Energy levels in different spin sectors and various fillings. The dielectric constant is fixed as 
		$\epsilon_r=15$. $E(S_z)$ are shown relative to the FM$_z$ state, and the results indicate that the FM$_z$ state is energetically 
		favorable for $\nu \gtrsim 0.2$.
	}
	\label{Fig_EnergyCell-eps15}
\end{figure}

Figure~\ref{Fig_EnergyCell-eps15} displays the relative energy $E(S_z) - E({\mathrm{FM}_z})$ 
per iDMRG unit cell, with respect to the fully polarized FM$_z$ sector (maximum $|S_z|$). Our findings reveal a finite gap above 
the FM$_z$ ground state for $\nu \gtrsim 0.2$, suggesting that the FM$_z$ order remains stable except for filling fractions $\nu$ very close to zero. Moreover, we simulate the model with $10 \leq \epsilon_r \leq 20$, with the results summarized in the magnetic phase diagram Fig.~1\textbf{d} in the main article.

\subsection{Ground-state results for $\nu=1$}
\begin{figure}[!htbp]
	\includegraphics[width=0.6\linewidth]{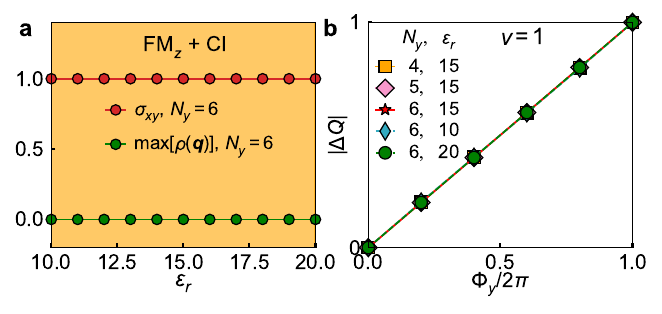} 
	\caption{ Uniform Chern insulator at $\nu=1$ filling. \textbf{a} Phase diagram containing only the CI state tuned by $\epsilon_r$, showcasing the maximum of $\rho({\boldsymbol{q}})$  and Hall conductivity $\sigma_{xy}$. \textbf{b} Charge pumping results $|\Delta Q|$ for various $N_y$ and $\epsilon_r$ parameters, measured after inserting a flux $\Phi_y=2\pi$. The inset provides a schematic illustration of the charge pumping simulation process.}
	\label{Fig_ICI}
\end{figure}

We present results of the model Eq.~(1) in the main text at an integer filling factor of $\nu=1$. By analyzing the ground-state energies across various spin-$S_z$ sectors, we consistently find that the ground states are always in the FM$_z$ ordered phase for 
dielectric constants $\epsilon_r \in [10,20]$, and for system sizes $N_y = 4, 5, 6$. Meanwhile, as shown in Fig.~\ref{Fig_ICI}\textbf{a}, the maximum $\rho({\boldsymbol{q}})\simeq 0$  in the whole range, indicating a uniform state. 

In the absence of interactions, the model Eq.~(1) can be viewed as a generalization of the renowned Haldane 
model \cite{Haldane1988Model}, which possesses a CI state with Chern number $|\mathcal{C}|=1$ and quantum Hall conductivity 
$\sigma_{xy}=1$. With interactions switched on, we simulate the Hall conductivity through charge pumping, as demonstrated in Fig.~\ref{Fig_ICI}\textbf{b} for several $\epsilon_r$ and $N_y=4, 5, 6$, with DMRG calculations. Our results reveal 
that the insertion of a single quantum flux ($\Phi_y=2\pi$) leads to the a charge pumping of $|\Delta Q|=1$, confirming that the ground 
state is indeed a CI state with a quantized Hall conductivity $\sigma_{xy}=\frac{2\pi|\Delta Q|}{\Phi_y} = 1$, even with the strong  coupling $U$ and $V(r)$. 

As shown in Fig.~\ref{Fig_ICI}\textbf{a}, through a scan of the parameter space, we identify a robust CI phase at least within the range $10 \leq \epsilon_r \leq 20$, indicating that the CI phase is highly stable against interactions and potentially persists over a wide range of dielectric constants. Nonetheless, in general cases the incorporation of interaction terms $U$ and $V(r)$ can profoundly enrich the electronic states, thereby requiring precision many-body calculations to determine their properties.

\begin{figure*}[!bp]
	\includegraphics[width=0.9\linewidth]{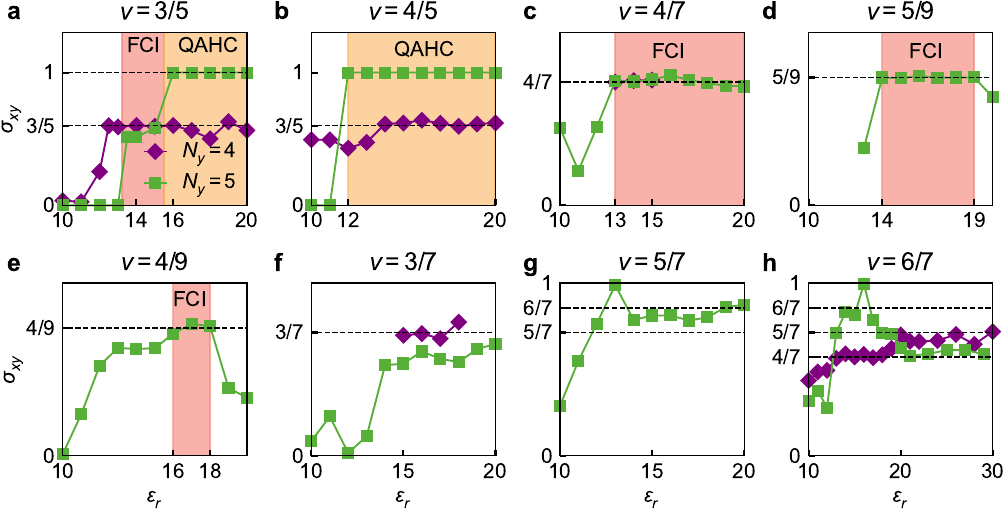} 
	\caption{Panels \textbf{a}-\textbf{h} show charge pumping results with varying $\epsilon_r$ for $\nu=3/5$, $4/5$, $4/7$, $5/9$, $4/9$, $3/7$, 
		$5/7$, and $6/7$, respectively. The presence of FCI and QAHC states on $N_y=5$ cylinders are marked with red and orange, 
		respectively. It is FM$_z$ for all these parameters. In all panels, the purple diamonds represent $N_y=4$ data, and the green 
		squares represent the $N_y=5$ results.
	}
	\label{SM_Fig_pump_phi}
\end{figure*}

\subsection{Ground-state results for fillings $\nu = p/q$ with $q\ge5$ } 
Besides the fillings $\nu = p/q$ with small denominators $q = 1, 2, 3$ discussed in the main text, we have also simulated larger values of $q = 5, 7, 9$. 

\subsubsection{Charge pumping results}
{\label{SM_subsec_chargepump}}  
As shown in Fig.~\ref{SM_Fig_pump_phi}, we present results of charge pumping simulations for various fractional fillings $\nu=p/q=3/5$, $4/5$, $3/7$, $4/7$, $5/7$, $6/7$, $4/9$, and $5/9$. As the denominator $q$ increases, the required XC length $N_x$ (which is typically several integer multiples of $q$) also increases, while the charge gap for potential FCIs may decrease. The simulations are more challenging and time-consuming, as smaller evolution step sizes and many more steps are needed to ensure adiabaticity. In practice, we limit $N_y$ to a maximum of 5 in our simulations and present results satisfying the adiabatic evolution condition. Further studies involving wider cylinders are necessary to validate our findings and better approach the thermodynamic limit.

Figures~\ref{SM_Fig_pump_phi}\textbf{c} and \textbf{d} highlight robust FCIs within a wide range, $13 \lesssim \epsilon_r \lesssim 20$ and 
$14 \lesssim \epsilon_r \lesssim 19$, for $\nu = 4/7$ and $5/9$, respectively. Panels \textbf{a} and \textbf{e} depict weaker FCIs with a narrower 
range ($\Delta\epsilon_r \sim 2$) for $\nu = 3/5$ and $4/9$. For the case of $\nu = 3/5$, a QAHC phase is observed for 
$16 \lesssim \epsilon_r \lesssim 20$ on the $N_y = 5$ cylinder. Distinctly, FCI states with $\sigma_{xy} \simeq \nu = 3/5$ emerges 
for $ N_y = 4$ cylinder. This difference is possibly due to geometric constraints, as discussed in the main text for the $\nu=1/2, 2/3$ 
cases. A similar phenomenon is observed for $\nu = 4/5$ in Fig.~\ref{SM_Fig_pump_phi}\textbf{b}, with a subtle difference that
$\sigma_{xy} = 3/5 \neq \nu$ for the FCI. 

For the case of $\nu =6/7$, we can observe results such as $\sigma_{xy}=4/7\neq\nu$ in Fig.\ref{SM_Fig_pump_phi}\textbf{h}. 
However, the relevant range of $\epsilon_r$ varies significantly with $N_y$, which suggests that further verification through simulations 
on larger system sizes is required. For the cases of $\nu=3/7$ and $\nu = 5/7$, although the Hall conductivity $\sigma_{xy}$ 
is found to be nonzero, the values are not (fractionally) quantized, indicating the absence of clear FCIs in Figs.~\ref{SM_Fig_pump_phi}\textbf{f} and \textbf{g}.

\begin{figure*}[!bp]
	\includegraphics[width=0.95\linewidth]{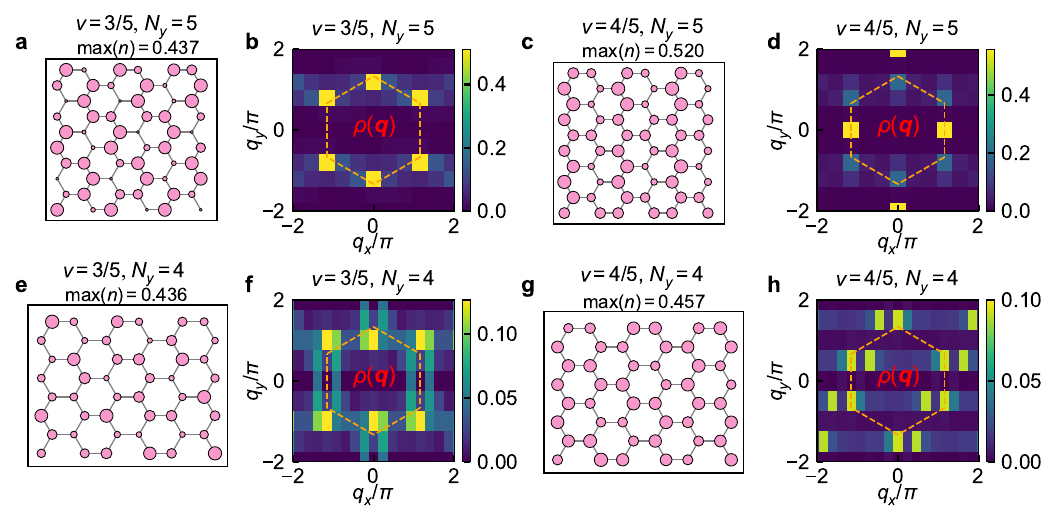} 
	\caption{DMRG results of \textbf{a} $n(\boldsymbol{r})$ and \textbf{b} $\rho(\boldsymbol{q})$ for QAHC with filling $\nu=3/5$ 
		(with $\epsilon_r=15$ and on $N_y=5$ cylinder). Panels (\textbf{e},\textbf{f}) show same quantities as panels (\textbf{a},\textbf{b}) but for FCI
		states on $N_y=4$. Panels (\textbf{c},\textbf{d},\textbf{g},\textbf{h}) show results for filling $\nu=4/5$ on $N_y=4, 5$ cylinders, with $\epsilon_r=18$. 
		The orange hexagon with dashed lines represent the first Brillouin zones of the honeycomb lattice. 
	}
	\label{SM_Fig_nu3545_D_eps_xi2_2by4}
\end{figure*}

For FCI states where $\sigma_{xy} < \nu$, such as for $\nu = 4/5$ (and possibly also for $\nu = 6/7$), we conjecture that this Hall conductivity might be attributed to the partial carriers involved in the formation of FCIs, while the remaining carriers form an insulating non-uniform charge distribution. As shown in $\rho(\boldsymbol{q})$ in Fig.~\ref{SM_Fig_nu3545_D_eps_xi2_2by4}\textbf{h}, these states exhibit peaks rather than vanished values, which is different from the typical FCI at $\nu = 1/3$ in Fig.~2\textbf{g}. Moreover, such states could also be classified as "fractal FCIs," arising from the full filling of fractal bands in the composite fermion picture \cite{lu2024fractional}.

\subsubsection{Density distribution for $\nu=3/5$ and $4/5$} 
In Fig.~\ref{SM_Fig_nu3545_D_eps_xi2_2by4}, we present the charge distribution $n(\boldsymbol{r})$ in real space and 
$\rho(\boldsymbol{q})$ in momentum space for $\nu = 3/5$ and $\nu = 4/5$ fillings. For $N_y = 5$, the charge distribution exhibits 
a well-defined CO structure for both fillings, which is consistent with the QAHC phase with Hall conductivity $\sigma_{xy} = 1$, as 
previously discussed. However, for $N_y = 4$, such a CO structure is incompatible with the limited width and thus disrupted, 
with its strength (i.e., the maximum of $\rho(\boldsymbol{q})$) becoming much weaker, and instead, a FCI state emerges, 
highlighting the sensitivity of the system to lattice geometry and the interplay between CO and topological orders.

\subsubsection{Phase diagram }  
\begin{figure}[!h]
	\centering
	\includegraphics[width=0.52\linewidth]{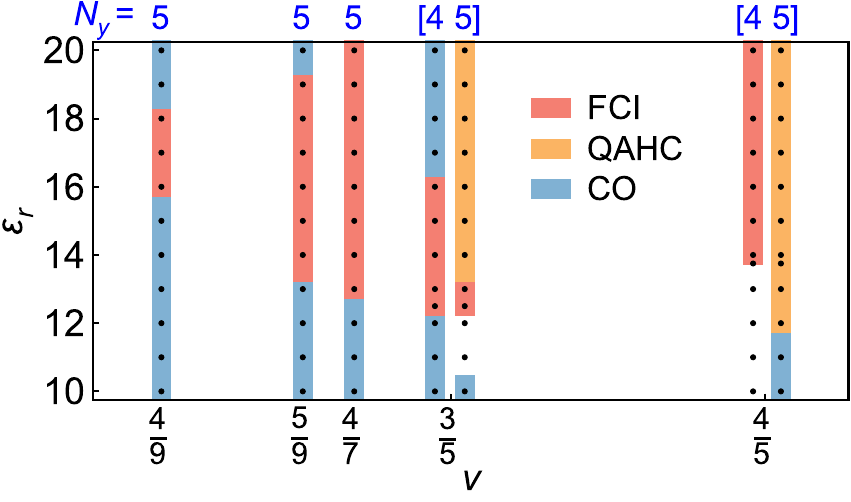} 
	\caption{The phase diagram with various fillings $ \nu = p/q $ for $ q \geq 5 $. Solid dots without color bar represent topologically 
		trivial states with uniform density distributions. The lattice widths used, $ N_y = 4 $ and $ 5 $, are indicated. For fillings $\nu = 3/5$ 
		and $ 4/5 $, both widths are considered. Across the entire phase diagram, FM$_z$ order is checked to be present.
	}
	\label{Other_FCI}
\end{figure} 

Above results for various fillings $ \nu = p/q $ withf $ q \geq 5 $ are summarized in the phase diagram~\ref{Other_FCI}. We observe robust FCIs for $\nu = 4/7$ (with $13 \leq \epsilon_r \leq 20$) and $\nu = 5/9$ ($14 \leq \epsilon_r 
\leq 19$), spanning a wide range of $\epsilon_r$. Additionally, FCIs with a narrower $\epsilon_r$ range ($\Delta \epsilon_r \sim 2$) 
are observed for $ \nu = 4/9$ and $3/5$. For $\nu =3/5$ and $4/5$ fillings, QAHC states are found for $N_y=5$. On the other hand, 
FCI states with $\sigma_{xy} = 3/5$ (for both $\nu =3/5$ and $4/5$ fillings) are found for $N_y = 4$ (see Figs.~\ref{SM_Fig_pump_phi} and \ref{SM_Fig_nu3545_D_eps_xi2_2by4} for details). The sensitivity of FCI or QAHC phase to the cylinder width is similar to that observed in the $\nu=1/2$ case discussed in the main text. Except for the topological states for $\nu=4/5$ filling that are not yet seen in experiments, the FCIs at $\nu = 3/5$, $4/7$, $4/9$, and $5/9$ obtained in our simulations are consistent with recent experiments. \\

\subsection{Phase diagram for incommensurate filling $\nu\simeq 0.63$ }
\begin{figure}[h]
	\includegraphics[width=0.55\linewidth]{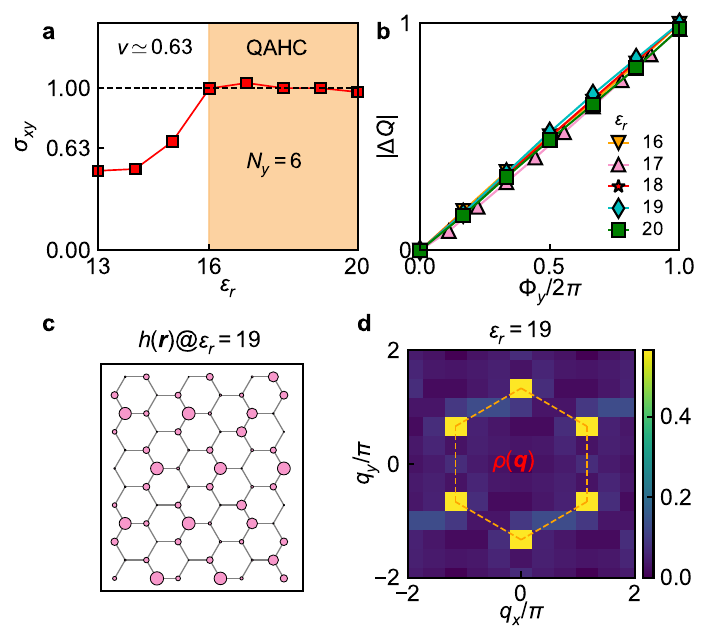} 
	\caption{  Emergence of QAHC states at incommensurate filling $\nu \simeq 0.63$ on a cylinder of width $N_y = 6$.
		\textbf{a}, Hall conductivity $\sigma_{xy}$ as a function of the dielectric constant $\epsilon_r$. The presence of QAHC states is highlighted in orange. All these states exhibit ferromagnetic polarization (FM$_z$).
		\textbf{b}, Charge pumping magnitude $|\Delta Q|$ for various QAHC states, measured after inserting a flux $\Phi_y = 2\pi$.
		\textbf{c}, Real-space hole density distribution $h(\boldsymbol{r})$, and
		\textbf{d}, corresponding charge structure factor $\rho(\boldsymbol{q})$ for the QAHC state at $\epsilon_r = 19$.
		The orange dashed hexagon denotes the first Brillouin zone of the underlying honeycomb lattice. Bond dimension $D=1200$ for all these simulations.  
	}
	\label{Fig_nu063_eps16_xi2_2by2_Tr}
\end{figure}

The discovery of QAHCs at incommensurate fillings is a fascinating and significant direction. We have already performed investigations at a filling of $\nu \simeq 0.63$, where such a crystal was recently reported in tMoTe$_2$~\cite{xu2025signature}. As shown in Fig.~\ref{Fig_nu063_eps16_xi2_2by2_Tr}, using iDMRG simulations at $\nu \simeq 0.63$ on a cylinder with $N_y = 6$ and varying $\epsilon_r \in [13, 20]$, we observed a QAHC of $\sigma_{xy} = 1$ for $\epsilon_r \in [16, 20]$ via charge pumping. Further analysis of the charge and hole structure reveals that the hole distribution $h(\boldsymbol{r})$ exhibits a pattern consistent with a crystal phase, and the static structure factor $\rho(\boldsymbol{q})$ shows peak structures similar to those at $\nu = 2/3$. This suggests that the electronic crystal at $\nu \simeq 0.63$ shares a similar origin with that at $\nu = 2/3$, as discussed in the main text.

Although our simulations have identified QAHCs at certain incommensurate fillings, establishing their existence over a continuous range of fillings \cite{lu2024extended} remains challenging. This is primarily due to the significantly larger unit cell sizes required to accurately model such incommensurate systems. Further numerical studies will be necessary to explore this more comprehensively, and we intend to pursue these in future work.

\section{Supplementary finite-temperature results}{\label{SM_FiniteT}}
\begin{figure}[!hbtp]
	\includegraphics[width=0.8\linewidth]{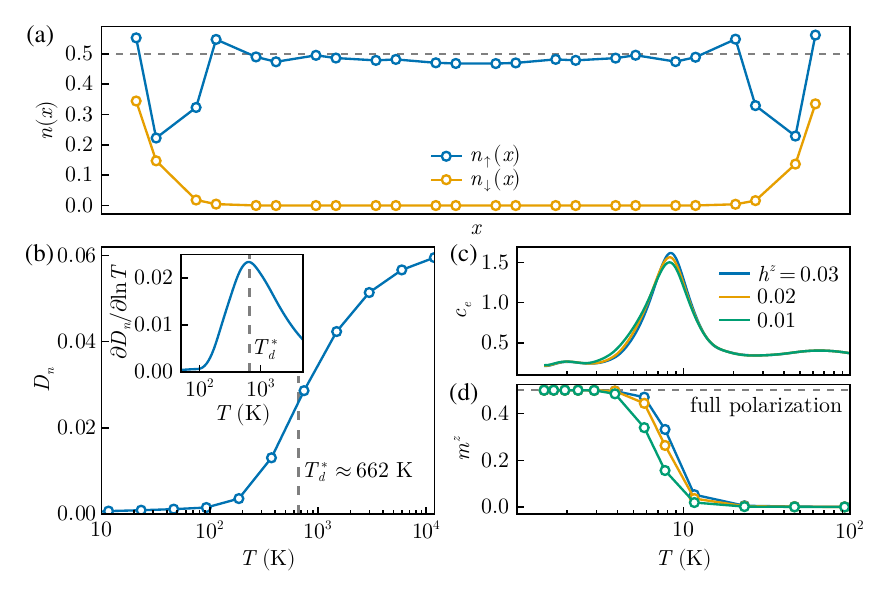} 
	\caption{Finite-temperature results for an XC$4\times12\times2$ 
		with parameters  $\nu=1$ and $\epsilon_r = 15$. (a) Electron density distribution $n_{\uparrow(\downarrow)}(x)$ at 
		$T \approx 1.45\textrm{ K}$. (b) Temperature evolution of the bulk-averaged double occupancy $D_n$. The inset 
		shows $\partial D_n/\partial \ln T$, which determines a characteristic temperature $T_d^\ast \approx 662$ K where $D_n$ 
		is suppressed rapidly. (c,d) Comparison of low-temperature electron specific heat $c_e$ and magnetic moment $m^z$ 
		per electron for different pinning fields $h^z$. Panels (a,b) and (c,d) are computed with bond dimension $D = 2000$ 
		and $D = 3000$, respectively.
	}
	\label{SM_Fig_Mz}
\end{figure}

In Fig.~\ref{SM_Fig_Mz}, we present additional finite-temperature benchmark results to validate the model simplification 
in the main text. It involves the reduction from local Hilbert space dimension $d = 4$ in the Hubbard-type model Eq.~(1)
to $d = 3$ in the spinful $t$-$V$ model Eq.~(2). The double-occupancy states of the spinful $t$-$V$ model are projected out. 
Since the case with filling $\nu = 1$ may potentially have the largest double occupancy $D_n$, we select $\nu = 1$ as a 
stringent test case.

Figure~\ref{SM_Fig_Mz}(a) shows the electron density distribution $n_\sigma(x) = \frac{1}{N_y} \sum_{i = 1}^{N_y} \langle 
n_{i\sigma}\rangle$, where $i$ runs over the $N_y$ sites sharing the same $x$-coordinates. Due to the open boundary along 
the $x$-direction, short-range oscillations appear. Thus, we use only the central 1/4 to 3/4 of the sites as bulk sites for averaging. 
This method is applied throughout the text below unless otherwise specified.

We calculate the average double occupancy $D_n = \frac{1}{N_{\rm bulk}} \sum_{i \in {\rm bulk}} \langle n_{i\uparrow} n_{i\downarrow} 
\rangle$ of the original model Eq.~(1) with local dimension $d = 4$. As shown in Fig.~\ref{SM_Fig_Mz}(b), $D_n$ gets suppressed at
about $T_d^\ast \approx 662$~K and almost diminishes as the temperature drops below about 100 K. Therefore, we conclude that  
the projected local Hilbert space dimension $d = 3$ in Eq.~(2) does not affect the low-temperature behavior of the original model Eq.~(1).

We also examine the impacts of the Zeeman pinning field. As depicted in Fig.~\ref{SM_Fig_Mz}(c, d), higher pinning fields $h^z$ 
marginally sharpen the FM phase transition but only slightly modify the transition temperature $T_c$. Consequently, our selection 
of $h^z = 0.01$ in the main text is thoroughly justified.

\section{computation method and convergence check of the spectral function}{\label{SM_Sec_Spectral}}
Let $\alpha\in \{A, B\}$ be the index of sublattice, fermionic operators in momentum space are defined as 
\begin{equation}
	c_{\bold{k}\alpha} = \frac{1}{\sqrt{N_xN_y}}\sum_{i \in \alpha} e^{-i\bold{k}\cdot \bold{r}_i}c_{i}.
\end{equation}  
The sublattice-resolved retarded Green's function reads
\begin{equation}
	\begin{aligned}
		G_{\alpha\beta}^{R}(\bold{k},\omega) =& -i\int_{0}^\infty dt e^{i\omega t} \expval{\left\{c_{\bold{k}\alpha}(t), c_{\bold{k}\beta}^\dagger\right\}}{\rm GS}\\ 
		=& -i\int_{0}^\infty dt e^{i\omega t} \left[\expval{c_{\bold{k}\alpha}(t)c_{\bold{k}\beta}^\dagger}{\rm GS} + \overline{\expval{c_{\bold{k}\alpha}^\dagger(t)c_{\bold{k}\beta}}{\rm GS}}\right]\\
		=& -i\int_{0}^\infty dt e^{i(\omega + E_g)t}\expval{c_{\bold{k}\alpha} e^{-iH t}c_{\bold{k}\beta}^\dagger}{\rm GS} + \overline{i\int_{0}^\infty dt e^{-i(\omega - E_g)t}\expval{c_{\bold{k}\alpha}^\dagger e^{-iH t}c_{\bold{k}\beta}}{\rm GS}}
	\end{aligned}
	\label{Eq:GR_komega}
\end{equation}
where $\ket{\rm GS}$ denotes the ground state.

It remains to compute the expectation values like $\expval{c_{\bold{k}\alpha} e^{-iH t}c_{\bold{k}\beta}^\dagger}{\rm GS}$ and $\expval{c_{\bold{k}\alpha}^\dagger e^{-iH t}c_{\bold{k}\beta}}{\rm GS}$.  Considering the former one, it can be written as the inner product of two states $e^{iHt/2}c_{\bold{k}\alpha}^\dagger\ket{\rm GS}$ and $e^{-iHt/2}c_{\bold{k}\beta}^\dagger\ket{\rm GS}$. We first use standard DMRG algorithm to obtain the ground state $\ket{\rm GS}$ that is represented as a MPS. Then, we act the operator $c_{\bold{k}\alpha}^\dagger$, which can be efficiently represented as a MPO with bond dimension $D = 2$ via automata, on the ground state MPS and obtain the new MPS $c_{\bold{k}\alpha}^\dagger\ket{\rm GS}$. $c_{\bold{k}\beta}^\dagger\ket{\rm GS}$ is obtained similarly. Then, we use the TDVP approach (accelerated with the controlled bond expansion) to perform the backward and forward time evolution to the former and the latter, which results in the two states $e^{iHt/2}c_{\bold{k}\alpha}^\dagger\ket{\rm GS}$ and $e^{-iHt/2}c_{\bold{k}\beta}^\dagger\ket{\rm GS}$, respectively. The inner product is calculated after each time step and the sequence of $\expval{c_{\bold{k}\alpha} e^{-iH t}c_{\bold{k}\beta}^\dagger}{\rm GS}$ is collected to be used in the numerical integration latter. Regarding $\expval{c_{\bold{k}\alpha} e^{-iH t}c_{\bold{k}\beta}^\dagger}{\rm GS}$ in the second term of Eq.~(\ref{Eq:GR_komega}), the procedure is similar except that the creation and annihilation operators are exchanged.  

\begin{figure*}[!tbp]
	\includegraphics[width=0.9\linewidth]{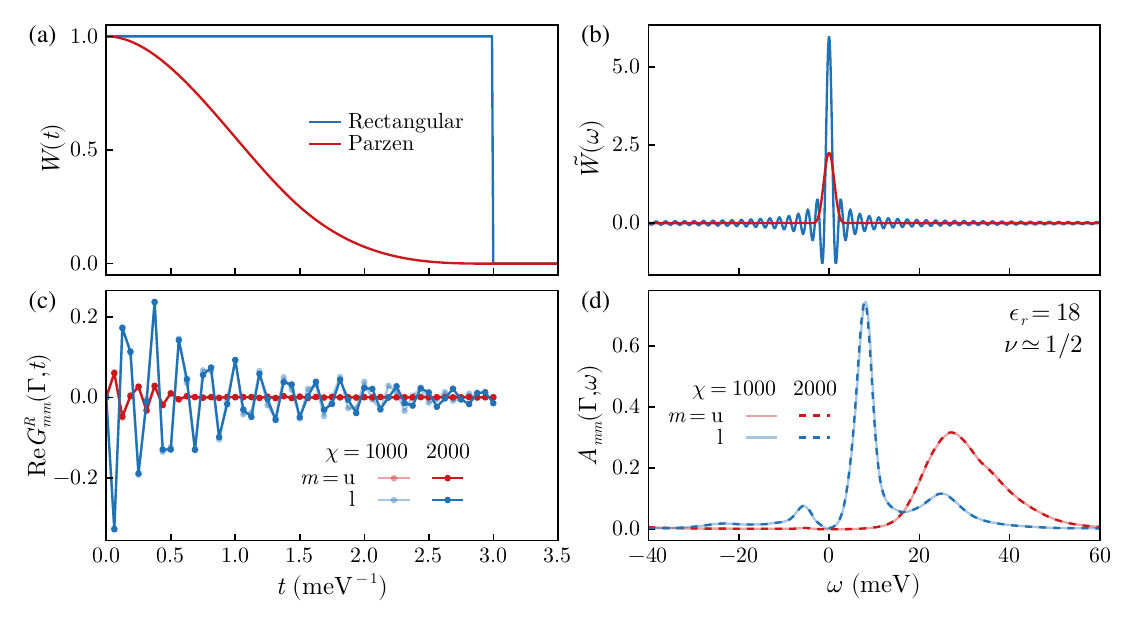} 
	\caption{(a) The rectangular window and Parzen window with $t_{\rm max} = 3$ meV$^{-1}$, whose Fourier transformations are shown
		in (b). (c) Data convergence of the real part of the retarded Green's function $\textrm{Re}G_{mm}^R(\textbf{k}, t)$ and (d) the corresponding 
		spectral function $A_{mm}(\textbf{k}, \omega)$. We show the $A_{mm}(\textbf{k}, \omega)$ results at $\textbf{k} = \Gamma$, evaluated 
		in the QAHC phase with $\nu \simeq 1/2$ and $\epsilon_r = 18$. The results provide representative benchmarks, since the presented
		spectral function in (d) exhibits several characteristic features, including a sharp peak, satellite peaks, and excitation continua.
	}
	\label{SM_Dynamics}
\end{figure*}

In the non-interacting case, the two bare bands are obtained by forming proper linear combinations of the fermionic operators from the two sublattices. A similar procedure can be applied in the interacting case by numerically diagonalizing the equal-time correlation matrices, i.e. 
\begin{equation}{\label{SM_Eq_diag1}}
	\left[
	\renewcommand{\arraystretch}{2}
	\begin{matrix}
		\expval{c_{\bold{k}A} c_{\bold{k}A}^\dagger} & \expval{c_{\bold{k}A} c_{\bold{k}B}^\dagger} \\
		\expval{c_{\bold{k}B} c_{\bold{k}A}^\dagger} & \expval{c_{\bold{k}B} c_{\bold{k}B}^\dagger}
	\end{matrix}
	\renewcommand{\arraystretch}{1}
	\right]  
	= V{\rm diag}\left[\expval{f_{\bold{k}{\rm l}} f_{\bold{k}{\rm l}}^\dagger}, \expval{f_{\bold{k}{\rm u}} f_{\bold{k}{\rm u}}^\dagger}\right]V^\dagger
\end{equation} 
where $f_{\bold{k}{\rm l}}$ and $f_{\bold{k}{\rm u}}$ is the fermionic operators belonging to the lower and upper bands, respectively, with the unitary matrix $V$ storing the coefficients of the linear combination. Written in the new representation, we get the band-resolved Green's function via $G_{mp}^R(\bold{k}, \omega) = \sum_{\alpha\beta} V_{\alpha m}^\ast G_{\alpha\beta}^R(\bold{k}, \omega) V_{\beta p}$ where $m, p \in \{{\rm l}, {\rm u}\}$ label the bands. Finally, computing spectral function $A_{mm}(\bold{k},\omega) = -2{\rm Im}G_{mm}^{R}(\bold{k},\omega)$ is straightforward after we get the retarded Green's function.

Due to the finite evolution time ($t_{\rm max} = 3$ meV$^{-1}$ in practical calculations), the numerical integration can only be performed 
after being multiplied by a window function $W(t)$ with vanished value beyond $t_{\rm max}$, i.e. 
\begin{equation}
	\begin{aligned}
		\widetilde{G}_{\alpha\beta}^{R}(\bold{k},\omega) \equiv& -i\int_{0}^{t_{\rm max}} dtW(t)e^{i(\omega + E_g)t}\expval{c_{\bold{k}\alpha} e^{-iH t}c_{\bold{k}\beta}^\dagger}{\rm GS} + \overline{i\int_{0}^{t_{\rm max}} dtW(t)e^{-i(\omega - E_g)t}\expval{c_{\bold{k}\alpha}^\dagger e^{-iH t}c_{\bold{k}\beta}}{\rm GS}}\\
		=& \int_{-\infty}^{\infty} \frac{d\omega^\prime}{2\pi} \widetilde{W}(\omega - \omega^\prime)G_{\alpha\beta}^{R}(\bold{k},\omega^\prime) \equiv \frac{1}{2\pi}\widetilde{W}(\omega) \ast G_{\alpha\beta}^{R}(\bold{k},\omega),
	\end{aligned}
\end{equation}
where $\widetilde{W}(\omega) \equiv \int_{-\infty}^{\infty} dt \, e^{i\omega t} W(t)$. Therefore, the effect of a finite $t_{\rm max}$ reduces to 
the convolution of a window function with the original Green's function. Directly performing the integration in $[0, t_{\rm max}]$ is equivalent 
to applying a rectangular window function, which leads to strongly non-physical oscillations in frequency domain. In practice, we use the 
Parzen window function
\begin{equation}
	W(t) = \left\{
	\begin{aligned}
		&1 - 6\abs{t/t_{\rm max}}^2 + 6\abs{t/t_{\rm max}}^3, &\abs{t} \leq t_{\rm max}/2\\
		&2(1 - \abs{t/t_{\rm max}})^3, &t_{\rm max}/2 < \abs{t} \leq t_{\rm max}\\
		&0, &\abs{t} > t_{\rm max},
	\end{aligned}		
	\right.
\end{equation}
with $\widetilde{W}(\omega) = 192\sin^4(\omega t_{\rm max}/4)/(\omega^4t_{\rm max}^3)$, as shown in Fig.~\ref{SM_Dynamics}(a,b).
The frequency resolution is about $8/t_{\rm max} \approx 2.67$ meV, defined by the full width at half maximum (FWHM) of the 
convolution kernel.

The single-particle spectral function data are obtained with a bond dimension $\chi = 1000$. The convergence with respect to the bond 
dimension for a typical QAHC case is shown in Fig.~\ref{SM_Dynamics}(c,d). Compared with a larger $\chi = 2000$, the difference in 
retarded Green's function $G_{mm}^R(\textbf{k},t)$ is very small during the time evolution, while the corresponding spectral function 
$A_{mm}(\textbf{k}, \omega)$ virtually coincide. This can be understood as the deviations in $G_{mm}^R(\textbf{k},t)$ at relatively long 
time $t$, become suppressed by the window function.

\end{document}